\documentclass[usenatbib]{mn2e}
\usepackage{graphicx, txfonts, natbib}
\usepackage{epsf}
\usepackage{amssymb}
\usepackage{epstopdf}
\usepackage{longtable,lscape}
\newcommand{\msun}{\mbox{M$_{\odot}$}}

\DeclareMathAlphabet{\mathsc}{OT1}{cmr}{m}{sc}
\def\testbx{bx}%
\DeclareRobustCommand{\ion}[2]{%
\relax\ifmmode
\ifx\testbx\f@series
{\mathbf{#1\,\mathsc{#2}}}\else
{\mathrm{#1\,\mathsc{#2}}}\fi
\else\textup{#1\,{\mdseries\textsc{#2}}}%
\fi}

\newcommand{\ha} {\mbox{H$\alpha$}}

\newcommand{\Caii} {[\ion{Ca}{ii}]}

\newcommand{\Oii} {[\ion{O}{ii}]}

\newcommand{\Mgi} {\ion{Mg}{i}}

\newcommand{\gps}{\ensuremath{g_{\rm P1}}}
\newcommand{\rps}{\ensuremath{r_{\rm P1}}}
\newcommand{\ips}{\ensuremath{i_{\rm P1}}}
\newcommand{\zps}{\ensuremath{z_{\rm P1}}}
\newcommand{\yps}{\ensuremath{y_{\rm P1}}}

\newcommand{\grizy}{\emph{griz}\yps}
\newcommand{\PS}{\protect \hbox {PS1}}

\newcommand{\ap}{PS1-11ap}
\newcommand{\dam}{PTF12dam}
\newcommand{\xk}{SN2011ke}
\newcommand{\bi}{SN2007bi}
\newcommand{\gx}{SN2010gx}

\begin{document}

\title[\ap]
{The superluminous supernova PS1-11ap:  bridging the gap between low and high redshift}

\author[M. McCrum et al.]
  {M.~McCrum,$^1$\thanks{E-mail: mmccrum04@qub.ac.uk}
  S. J.~Smartt,$^1$
  R.~Kotak,$^1$ 
  A.~Rest,$^2$
  A.~Jerkstrand,$^1$
  C.~Inserra,$^1$
  S. A.~Rodney,$^{3,17}$
    \newauthor
     T.-W.~Chen,$^1$
  D. A.~Howell,$^{4,5}$
  M. E.~Huber,$^6$
  A.~Pastorello,$^7$
 J. L.~Tonry,$^6$ 
 F. ~Bresolin,$^6$ 
\newauthor
    R.-P.~Kudritzki,$^6$
  R.~Chornock,$^8$
  E.~Berger,$^8$
  K.~Smith,$^1$  
 M. T.~Botticella,$^9$ 
 R. J.~Foley,$^8$
 \newauthor
 M.~Fraser,$^1$
D.~Milisavljevic,$^8$ 
M.~Nicholl,$^1$
A. G.~Riess,$^3$
C. W.~Stubbs,$^8$
S.~Valenti,$^{4,5}$ 
\newauthor
W. M.~Wood-Vasey,$^{13}$ 
D.~Wright,$^1$
D. R.~Young,$^1$
M.~Drout,$^8$
I.~Czekala,$^8$
W. S.~Burgett,$^6$
\newauthor
K. C.~Chambers,$^6$
P.~Draper,$^{14}$
H.~Flewelling,$^6$
 K. W.~Hodapp,$^6$
 N.~Kaiser,$^6$ 
 E. A.~Magnier,$^6$
\newauthor
 N.~Metcalfe,$^{14}$
 P. A.~Price,$^{16}$
 W.~Sweeney,$^6$
 R. J.~Wainscoat$^6$ \\
  $^1$Astrophysics Research Centre, School of Maths and Physics, Queen's University Belfast, Belfast BT7 1NN, UK\\
  $^2$Space Telescope Science Institute, 3700 San Martin Drive, Baltimore, MD 21218, USA\\
  $^3$Department of Physics and Astronomy, Johns Hopkins University, 3400 North Charles Street, Baltimore, MD 21218, USA\\
  $^4$Las Cumbres Observatory Global Telescope Network, 6740 Cortona Dr., Suite 102, Goleta, California 93117, USA \\
  $^5$Department of Physics, University of California Santa Barbara, Santa Barbara, CA 93106, USA \\
  $^6$Institute for Astronomy, University of Hawaii at Manoa, Honolulu, HI 96822, USA\\
  $^7$ INAF - Osservatorio Astronomico di Padova, Vicolo dell`Osservatorio 5, 35122 Padova, Italy\\
  $^8$Department of Physics, Harvard University, Cambridge, MA 02138, USA\\
  $^9$INAF - Osservatorio astronomico di Capodimonte, Salita Moiariello 16, I- 80131 Napoli, Italy\\  
  $^{10}$INAF Osservatorio Astronomico di Brera, via Bianchi 46, I-23807 Merate, Italy\\
   $^{11}$ California Institute of Technology, 1200 E. California Blvd., Pasadena, CA 91125, USA\\
  $^{12}$Institute of Astronomy, National Central University, Chung-Li 32054, Taiwan\\
  $^{13}$PITT PACC, Department of Physics and Astronomy, University of Pittsburgh, Pittsburgh, PA 15260, USA\\  
  $^{14}$Department of Physics, Durham University, South Road, Durham DH1 3LE, UK\\
   $^{15}$US Naval Observatory, Flagstaff Station, Flagstaff, AZ 86001, USA\\ 
   $^{16}$ Department of Astrophysical Sciences, Princeton University, Princeton, NJ 08544, USA\\
   $^{17}$ Hubble Postdoctoral Fellow\\ \\
   $\star$ \emph{E-mail: mmccrum04@qub.ac.uk}\\
}
\maketitle

\begin{abstract}

  We present optical photometric and spectroscopic coverage of the
  superluminous supernova (SLSN) \ap, discovered with the Pan-STARRS1 Medium
  Deep Survey at $z=0.524$.  This intrinsically blue transient rose
  slowly to  reach a peak magnitude of $M_u=-21.4$ mag and bolometric
  luminosity of $8\times10^{43}$\,ergs$^{-1}$ before settling onto a
  relatively  shallow gradient of decline. The observed decline is
  significantly slower than those of the superluminous type Ic SNe which have
  been the focus of much recent attention. Spectroscopic similarities with
 the lower redshift  \bi\ and a decline rate similar to $^{56}$Co
 decay timescale initially indicated that this transient
 could be a candidate for a pair instability supernova (PISN)
 explosion.  Overall the transient appears
quite similar to \bi\ and the lower redshift object \dam. 
The extensive data set, from 30 days before peak to 
230 days after allows a detailed and quantitative comparison with
published models of PISN explosions.  We find that the \ap\ data
do not match these model explosion parameters well, supporting the
recent claim that these SNe are not pair instability explosions. 
We show that \ap\ has many features in common with the faster
declining superluminous Ic supernovae, and the lightcurve
evolution can also be quantitatively explained by the magnetar spin
down model. At a redshift of $z=0.524$ the observer frame optical coverage
provides comprehensive restframe UV data and allows us to compare it with the superluminous SNe recently found at high
redshifts between $z = 2 - 4$.   While these high-z explosions are
still plausible PISN candidates, they match the photometric evolution
of PS1-11ap and hence could be counterparts to this lower redshift
transient. 
\end{abstract}


\section{Introduction}
\label{intro}

Recent wide field surveys, such as the Palomar Transient Factory (PTF) \citep{ptfa,ptfb} and Pan-STARRS1 (PS1) \citep{PS1_system}
have uncovered that the schema for classifying the highly luminous and
energetic endpoints of massive stars known as supernovae (SNe) may be
more complicated than previously thought. \cite{quimb} linked the
peculiar SCP06F6 \citep{scp06f6} with a number of PTF discoveries to
suggest the emergence of a new class of superluminous supernovae (SLSNe)
characterised by peak absolute magnitudes of $<-21$ mag with total
radiated energies of $\sim10^{51}$ erg and a preference for low
luminosity hosts. A number of similar objects were discovered at
higher redshifts with PS1 cementing the notion that new types of SNe
were being discovered \citep{chom, berg, chorn, lun}. 

\cite{gal-yam} proposed a three tier classification scheme for these objects of
SLSNe-I, SLSNe-II and SLSNe-R based on the observed photometric,
spectroscopic and supposed physical properties. \cite{11xk} have recently
presented an extensive study of low redshift SLSNe with data before
peak and late photometric coverage to beyond 200 days. 
As found by 
\cite{10gx} for SN2010gx, these SNe evolve to resemble type Ic SNe but on much
slower timescales. Hence \cite{11xk} have termed this class 
SLSN-Ic as they are superluminous and (in the standard classification
definitions) of type Ic as they lack hydrogen and helium lines. These are
the type called SLSNe-I by \cite{gal-yam} (and elsewhere in the literature) 
but we will use the term SLSN-Ic throughout this paper. 
\cite{11xk} have proposed that the lightcurves for these SLSN-Ic can be 
quite well fit with the deposition of energy from a spinning down
magnetic neutron star. 

 Of particular interest here is the so-called SLSNe-R class which
 derives its name from the slow decline in the post-peak light curve.
 This luminosity could possibly be powered by the radioactive decay of
 a large mass of $^{56}$Ni created in the explosion decaying to
 $^{56}$Co and $^{56}$Fe, hence the `R' in the name. One event that
 has previously been classified as such is \bi\ \citep{07bia,07bib}. This SLSN
 was proposed by \cite{07bia} to be the result of electron-positron
 pair production in the core of an initially very massive
 ($M_{initial}>140 \msun$) star which reduces supporting radiation
 pressure and leads to gravitational contraction with a rise in core
 temperature above $10^9$K. The carbon-oxygen core at this point must
 be more massive than $\sim60\msun$ and undergoes explosive oxygen burning
 which can create an unusually large mass of $^{56}$Ni. This Pair
 Instability Supernova (PISN), and the subsequent decay of the
 $^{56}$Ni would provide the energy required to create the observed
 luminosity and unbind the star completely, leaving no remnant
 \citep{pisna,pisnb,pisnc,pisnd,2007A&A...475L..19L}.
 
 Another possible mechanism for producing the large amount of $^{56}$Ni required is the
 evolution of a slightly less massive carbon-oxygen core ($M<45
 \msun$) until the more usual iron core-collapse (CCSN) process
 \citep{massfea,massfeb}. This alternate process can reproduce the
 light curve shape of these objects but \cite{gypisn} argued that
 they should lead to stronger nebular emission lines of Fe than were
 detected in the late-time spectra of SN2007bi. \cite{cooke} proposed
 the discovery of another two such PISN candidates, thought to be at redshifts
 of $z\sim2$ and $z\sim4$, and PTF has discovered another possible
 candidate, PTF10nmn (Yaron et al, in prep).

However, the massive progenitors required for such a sequence of
events to take place require the low metallicity environments such as
those found only in the very early universe \citep{pisngal,pisngal2}.
Emission lines from the host spectra of \bi\ indicate that it is only
at $z=0.128$, and \cite{07bib} estimated a metallicity of around
0.3Z$_{\odot}$. This is much higher than what was originally expected
for PISN or massive CCSNe progenitors, as by definition a massive carbon-oxygen core must
have low enough mass-loss rates that it does not shrink to the typical
Wolf-Rayet (WR) stellar masses \citep[around
5-25M$_{\odot}$][]{2007ARA&A..45..177C} through strong radiatively
driven winds. There is an obvious problem for SN2007bi in that we lack
a physical model for the progenitor to exist with a core mass of
around 60$\msun$ at this metallicity, if the typical mass-loss rates
of massive stars at 0.3Z$_{\odot}$ or LMC type metallicities hold
\citep{2002A&A...392..653C,2007ARA&A..45..177C,2007A&A...473..603M}.
Stellar evolutionary models all end with carbon-oxygen cores of much
lower mass
\citep[e.g.][]{pisnd,2003ApJ...591..288H,2004MNRAS.353...87E,2007A&A...475L..19L}.
Although \cite{2007A&A...475L..19L} can produce PISN at metallicities
seen in the nearby Universe, these models all produce hydrogen rich
events and WR progenitor model explosions require
significantly lower metallicity, typically below $0.05$Z$_{\odot}$.
If, however, very massive stars ($M_{initial}\sim320 \msun$) exist, \cite{yusof} show that a PISN can be produced at $0.3$Z$_{\odot}$.
For metallicites such as in the Small Magellanic Cloud (SMC), \cite{yusof} manage to synthesise PISNe from stars with $100<M_{initial}<190 \msun$, where the very large convective cores of the stars during their main sequence phase allows them to evolve homogeneously.   To achieve this however, the mass loss rates of these models must be set so that the stars lose their H envelope but retain their He layer, predicting that the resulting SNe may be of Type Ib as supposed to the observed Ic-like properties seen in \bi.
\cite{2012MNRAS.424.2139D} show that the detection of the He lines for a Type Ib SN requires not only the presence of helium but also the excitation source of $^{56}$Ni to be well mixed.  Hence a Type Ic PISN from the \cite{yusof} models is plausible.

\begin{figure}
\includegraphics[scale=0.205]{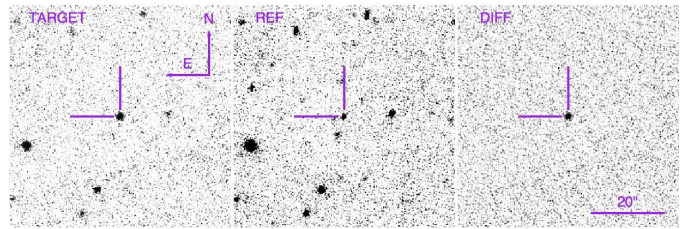}
\caption{\PS\ target, reference and difference images of \ap.  The images are \ips-band.}
\label{fig:descim}
\end{figure}

Another very luminous supernovae, \dam, has recently been discovered
at the similar redshift of $z=0.107$. \dam\ \citep{12dam} shares
similar photometric and spectroscopic properties to \bi\ but a
substantial dataset has allowed for more extensive modelling that
points away from the PISN or $^{56}$Ni-driven explanation. One possible
alternative is that the energy injection required to explain the
luminosity comes from the spin-down of a fast rotating neutron star or
magnetar, which boosts the normal SN luminosity generating mechanisms.
\cite{11xk} have proposed that this mechanism could explain the faster
declining SLSN-Ic in their low redshift sample, based on the recent
theoretical treatment of \cite{magmod} and \cite{2010ApJ...719L.204W}.
The idea of neutron star remnants powering the subsequent SN evolution
dates back to \cite{pulsara} and the energy input of highly magnetic
neutron stars has been proposed to influence gamma-ray burst
production \citep{1992Natur.357..472U,2000ApJ...537..810W}. The
production of hyper-energetic SNe due to the extraction of energy from
a spinning down magnetar was suggested by \cite{pulsarb}, providing
means for producing SNe with total radiated energies greater than
$10^{51}$\,ergs.  In this paper we apply similar analysis to the SLSN
\ap\ as done in \cite{11xk} and \cite{12dam}. We 
 add fuel to the argument that if PISNe do exist, observational
evidence for them has not yet been found.

\section{Photometry}

\subsection{Pan-STARRS1}

The \PS\ system is a high-etendue wide-field imaging system,
designed for dedicated survey observations. The system is installed on
the peak of Haleakala on the island of Maui in the Hawaiian island
chain. The telescope has a 1.8-m diameter primary mirror and the
gigapixel camera (GPC1) located at the $f/4.4$ cassegrain focus
consists of of sixty 4800$\times$4800 pixel detectors (pixel scale
0.258'') giving a field of view of 3.3$^{\circ}$ diameter. Routine
observations are conducted remotely, from the Waiakoa Laboratory in
Pukalani. A more complete description of the \PS\ system, both
hardware and software, is provided by \cite{PS1_system}. The survey
philosophy and execution strategy are described in \cite{PS_MDRM}

The \PS\ observations are obtained through a set of five broadband
filters, which are designated as \gps, \rps, \ips, \zps, and \yps.
Although the filter system for \PS\ has much in common with that used
in previous surveys, such as the Sloan Digital Sky Survey (SDSS)
\citep{SDSS}, there are important differences. The \gps\ filter
extends 20~nm redward of $g_{SDSS}$, paying the price of 5577\AA\ sky
emission for greater sensitivity and lower systematics for photometric
redshifts, and the \zps\ filter is cut off at 930~nm, giving it a
different response than the detector response which defined
$z_{SDSS}$. SDSS has no corresponding \yps\ filter. Further
information on the passband shapes is described in \cite{PS_lasercal}.
The \PS\ photometric system and its response is covered in
detail in \cite{PS1phot}. Photometry is in the ``natural'' \PS\
system, $m=-2.5\log(flux)+m'$, with a single zeropoint adjustment $m'$
made in each band to conform to the AB magnitude scale.

This paper uses images and photometry from the \PS\ Medium-Deep
Field survey (MDS), the strategy of which are described in \cite{JTwds}.
Observations of between 3-5 medium deep (MD) fields are taken each night
and the filters are cycled through in the following pattern : \gps\ and
\rps\ in the same night (dark time), followed by \ips\ and \zps\ on the
subsequent second and third nights respectively. Around full moon only
\yps\ data are taken. Any one epoch consists of 8 dithered exposures of
either $8\times113$s for \gps\ and \rps\ or $8\times240$s for the other
three, giving nightly stacked images of 904s and 1632s duration.

Images obtained by the \PS\ system are processed through the
Image Processing Pipeline (IPP) \citep{PS1_IPP}, on a computer cluster
at the Maui High Performance Computer Center (MHPCC). The pipeline runs the
images through a succession of stages including device
``de-trending'', a flux-conserving warping to a sky-based image plane,
masking and artifact location. De-trending involves bias and dark
correction and flatfielding using white light flatfield images from a
dome screen, in combination with an illumination correction obtained
by rastering sources across the field of view. After determining an
initial astrometric solution the flat-fielded images were then warped
onto the tangent plane of the sky using a flux conserving algorithm.
The plate scale for the warped images was originally set at 0.200
arcsec/pixel, but has since been changed to 0.25 arcsec/pixel in what
is known internally as the V3 tesselation for the MD fields. Bad pixel
masks are applied to the individual images and carried through the
stacking stage to give the ``nightly stacks'' of 904s and 1632s total
duration. Two independent difference imaging pipelines run on a daily basis on
the MD fields. 

The \PS\ project has developed the Transient
Science Server (TSS) which takes the nightly stacks created by the IPP
in MHPCC, creates difference images with respect to high quality
reference images, and then carries out point source function (PSF) fitting photometry on the
difference images to produce catalogues of variables and transient
candidates \citep[initially described in][]{suvi}. All these
individual detections are ingested into a MySQL database residing at
Queen's University Belfast (QUB) after an initial rejection algorithm based
on the detection of saturated, masked or suspected defective pixels
within the PSF area. Sources are assimilated into transient candidates
if they have more than 3 quality detections within the last 7
observations of the field, including detections in more than one filter,
and an RMS scatter in the positions of $\leq0\farcs5$. Each quality
detection must be of more than $5\sigma$ significance (defined as
instrumental magnitude error $<0.2^{m}$) {\em and} have a Gaussian
morphology ($XY_{moments} < 1.2$). Transient candidates which pass
this automated filtering system are promoted for human screening,
which currently runs at around 10\% efficiency (i.e. 10\% of the
transients promoted automatically are judged to be real after human
screening). These
 real transient candidates are crossmatched with all available catalogues of 
astronomical sources in the MDS fields. We use our own MDS catalogue and also extensive 
external catalogues (e.g. SDSS, GSC, 2MASS, APM,
Veron AGN,  X-ray catalogues) in order to have a first pass
classification of supernova, variable star, active galactic nuclei or nuclear transient.

In parallel, an independent set of difference images are produced at the Centre for 
Astrophysics (Harvard) on the Odyssey cluster from these IPP nightly stacked images
using the $photpipe$ \citep{psfphot} software. A custom built reference stack is produced and 
each IPP nightly stack uses this to produce an independent difference image. 
This process is described in \cite{2010ApJ...720L..77G,chom,berg,suvi, chorn,lun}
and potential transients are visually
inspected for promotion to the status of transient
alerts. We cross match between the TSS and the $photpipe$ transient streams and 
agreement on the detection and photometry is now excellent, particularly after the 
application of uniform photometric calibration based on the ``ubercal'' process \citep{2012ApJ...756..158S,2013ApJS..205...20M}.

\subsection{Photometric observations}
\label{phot}

\begin{figure*}
 \includegraphics[angle=270, scale=0.6]{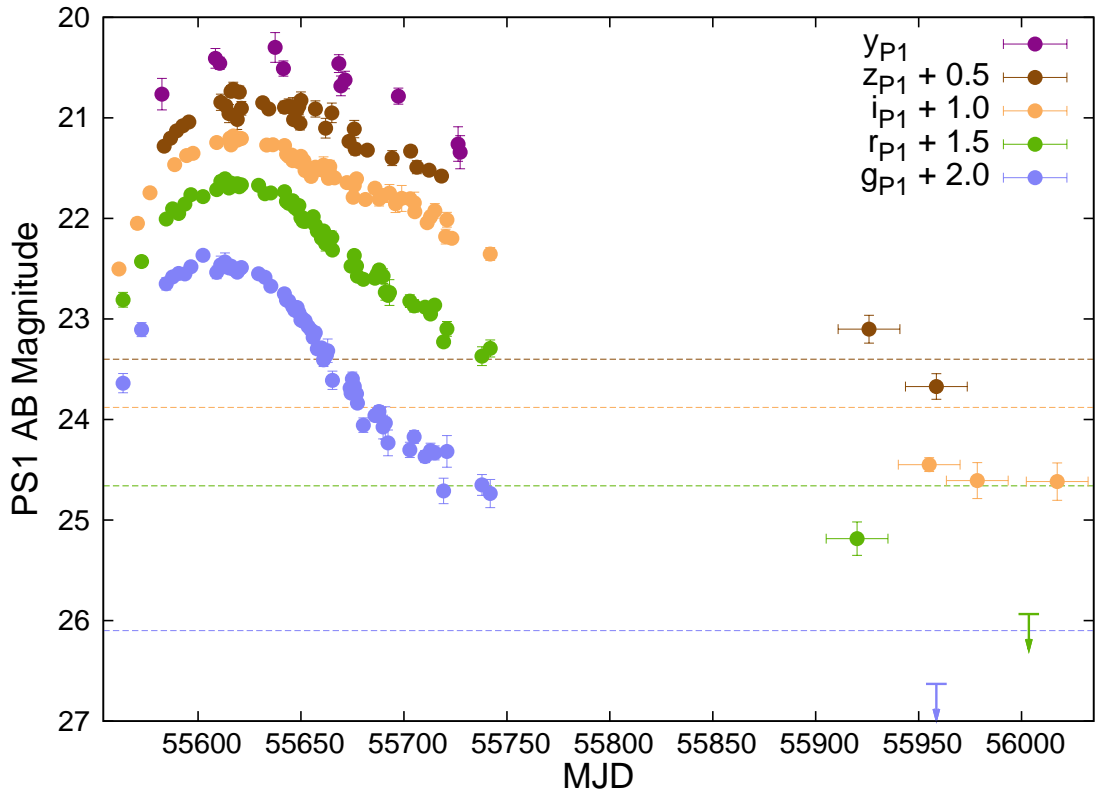}
 \caption{Optical photometry for \ap.  The dashed lines show the host galaxy magnitudes from the pre-explosion \PS\ images also with linear offsets as noted in the figure key and the arrows indicate limiting magnitudes when a detection of the transient could not be obtained above a 3$\sigma$ limit.}
\label{fig:11ap_lc}
\end{figure*}

\ap\ was discovered as a bright
transient object in the \PS\ Medium Deep Field 05 (MD05) on the 31st
December 2010 at a J2000 location of RA\,=\,10$^h$48$^m$27$^s$.73,
Dec\,=\,57$^\circ$09$'$09$''$.2.  As a high significance transient, 
it was discovered simultaneously in both the TSS and $photpipe$ parallel searches. 
Fig.\,\ref{fig:descim} shows that a faint host is present in the \PS\ reference
image but that the post explosion image shows the transient as
significantly brighter than its host galaxy. The object was initially
observable from Hawaii until the 29th June 2011, at which point it
disappeared behind the sun, and \grizy\ exposures were taken
throughout this entire period (referred to here as the first season).
Accompanying this was \emph{griz} photometry (SDSS filters) taken with the 2-m
Liverpool Telescope + RATCam (LT) on La Palma which was useful for
completeness when \PS\ was experiencing downtime from bad weather.
This additional photometry was particularly useful during the period
between the Modified Julian Dates (MJD) of 55610 and 55640 when the
supernova was at its peak brightness as \PS\ was unavailable at this
time. A MJD $=55613$ is used as the peak epoch of the SN, deduced by
fitting a low order polynomial to the photometric \rps-band data. \PS\ started
observing MD05 again on MJD $=55904$ (9th December 2011) and the
transient was still visible in the \rps, \ips\ and \zps-bands
(referred to here as the second season). A number of photometric
points could be retrieved at this date by manually subtracting
pre-explosion reference images from each epoch to remove any flux
contamination from the faint host galaxy and by co-adding multiple
exposures together, the methodology of which is subsequently
explained.

Photometry during the first season of \PS\ data was carried out using
difference images and measurement routines within the $photpipe$
pipeline \citep{psfphot} using stacked pre-explosion \PS\ exposures as
a reference image. Details of these photometric measurements are
presented in \cite{chom,berg,suvi} and 
 \cite{12dam} and will be given in more
detail in \cite{psfphot2} and \cite{psfphot3}. The LT data were processed through 
the LT detrending system which debiases and flat-fields images. We built our own 
 fringe frames to subtract from the \emph{i} and \emph{z}-band exposures. 
Photometric measurements of PS1-11ap were completed
by performing PSF-fitting photometry 
within \textsc{iraf}\footnote{\textsc{iraf} is distributed by the
  National Optical Astronomy Observatories, which are operated by the
  Association of Universities for Research in Astronomy, Inc., under
  the cooperative agreement with the National Science Foundation.}
using the custom built \emph{snoopy}\footnote{\emph{snoopy} was originally devised by F. Patat (1996) and implemented in \textsc{iraf} by E. Cappellaro.  The package is based on \emph{daophot} and has been optimised for SNe.} package.
No image subtraction was carried out on the LT data as the transient was significantly brighter than its host.
The effects of fringing were particularly
apparent in the \emph{z}-band exposures and were difficult to
completely remove at times hence the increased scatter in the LT
\emph{z}-band photometry. Zero points for each image were calculated
using magnitudes from a number of bright SDSS stars. Filter
transformations, as detailed in \cite{PS1phot}, were applied to each
filter to correct all the SDSS \emph{griz} magnitudes to \PS\
\emph{gri}\zps\ magnitudes. 

Difference imaging of the second season
data was carried out by creating custom reference images from
pre-explosion \PS\ exposures obtained during the period between MJD
55160 and 55350. Approximately 8 nightly stacks were manually picked
as high quality images in each filter and were median combined using the
\emph{imcombine} task in \textsc{iraf} for each filter. These
reference images could then be subtracted from the second season \PS\
observations using
\textsc{hotpants}\footnote{http://www.astro.washington.edu/users/becker/hotpants.html},
an implementation of the Alard algorithm \citep{isis}. This is the same algorithm 
as used in $photpipe$ for earlier epochs, hence the methods are consistent.
 To improve the
signal in the images the data were binned into 30 day periods and
PSF-fitting photometry was again performed using \emph{snoopy}. The
detections show a slowly fading tail phase during this late monitoring
phase and the co-added frames considerably help with detections.
During the 3 year lifetime of \PS\ no previous outbursts for \ap\ have
been detected before the first detection in December 2010.
The \grizy\ light curves can be
seen in Fig.\,\ref{fig:11ap_lc} where the dotted lines represent the
host magnitudes, obtained by performing aperture photometry on the
pre-explosion, combined images (see Sect.\,\ref{host}). 

\begin{table*}
 \caption{Central rest wavelengths ({\AA}) of optical passbands for each of the SLSNe used in this paper.  The numbers in \emph{italic} show the filters used for comparisons.  The redshift of each object is also given in the top row.}
 \label{tab:11ap_k}
 \begin{tabular}{@{}lccccccc}
  \hline
  \hline
Filter && \ap & \dam & \bi & \xk  & \gx & SN2213-1745 \\
\hline
& - & 0.524 & 0.107 & 0.127 & 0.143 & 0.230 & 2.05 \\
\hline
u & 3540 & - & \emph{3196} & - & 3097 & 2878 & - \\
B & 4445 & - & - & 3945 & 3890 & - & - \\
g & 4860 & \emph{3193} & \emph{4396} & - & 4257  & 3878 & $\sim1600$ \\
V & 5505 & - & - & 4885 & 4817 & -  & - \\
r  & 6230 & 4091 & \emph{5632} &  \emph{5528} & \emph{5455} & 5065 & $\sim2050$\\
i & 7525 & 4939 & 6799 & 6673 & 6585 & \emph{6199} & \emph{$\sim$2530}\\
z  & 8660 & \emph{5680} & 7820 & - & 7573 & 7426 & - \\
y  & 9720 & 6382 & 8786 & - & - & - & - \\
\hline
 \end{tabular}
 \medskip
\end{table*}

\begin{figure*}$
\begin{array}{cc}
\includegraphics[angle=270, scale=0.425]{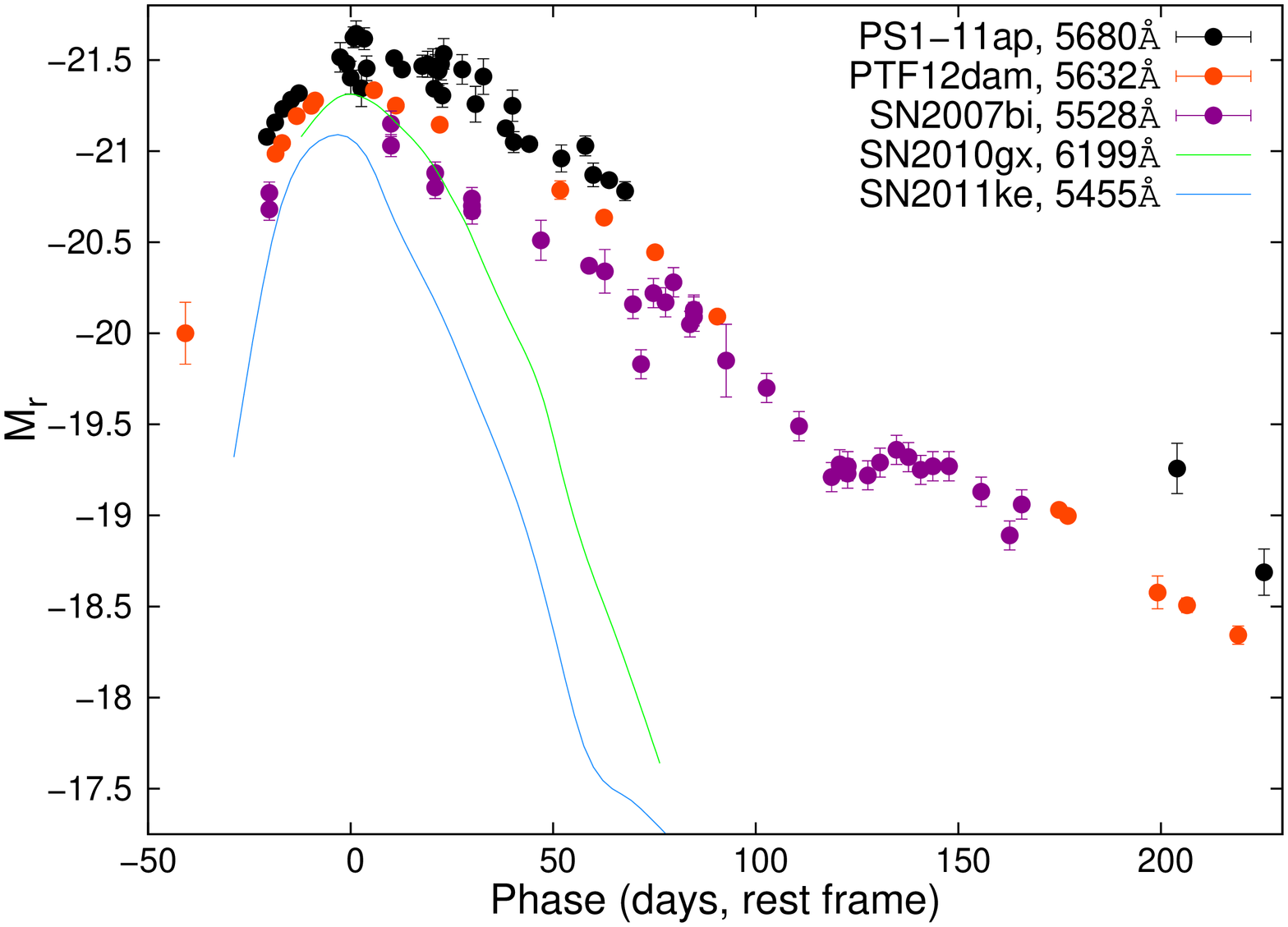} \\
\includegraphics[angle=270, scale=0.425]{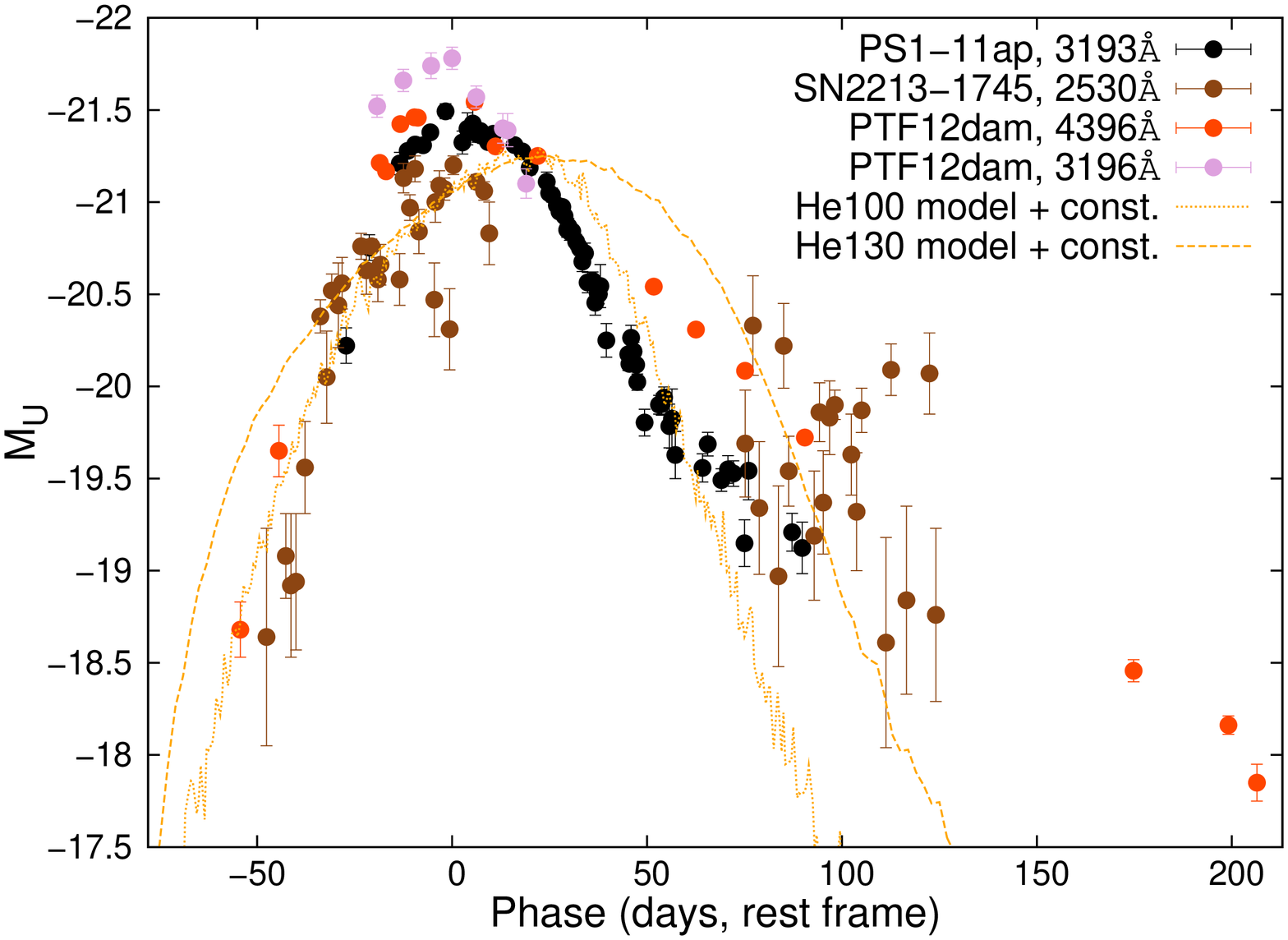} \\
\end{array}$
\caption{\emph{(TOP PANEL)} An absolute magnitude comparison of \zps
  -band \ap\ data with \emph{r}-band \dam, \bi\, \xk\ and
  \emph{i}-band \gx\ data.  See text for appropriate references. All
  of these are absolute AB magnitudes calculated without a detailed
  $K-$correction using Eq. 1. 
 \emph{(BOTTOM PANEL)}  A comparison of \gps -band \ap\ with \emph{i}-band SN2213-1745 and \emph{g} and \emph{u}-band \dam.  Also included here are two PISNe models similar to that used by Cooke \emph{et al.} (2012) to fit the SN2213-1745 data.  As in the Cooke \emph{et al.} (2012) paper, the models have been offset to match the luminosity of the data.}
 \label{fig:11apab}
\end{figure*}

The two arrows during the season 2
data points indicate epochs at which no transient could be detected at
the position of \ap\ at greater than three times the level of the
background noise.
Of note regarding the colour evolution of the light curves is that the \gps\ and \rps-band light curves decline much quicker than the redder filters.  This is to be expected as heavier elements in the spectra for \ap\ at the derived rest wavelength covering the \gps-band filter may cause a line blanketing effect.  This is discussed in Section\,\ref{spectra} along with a more in-depth look at the temperature evolution of \ap.  

K-correction values were calculated using spectra from the 22nd February, 11th March, 22nd April and 22nd June. 
As these observations were reasonably frequent no intermediate
measures were required to fill in lengthly gaps between epochs when a
spectrum was not available. All the photometric values deduced for
\ap\ could then be corrected at all epochs using the nearest possible
\emph{K}-correction value.
At the redshift of \ap\ the conversion filters correspond to the
following rest frame filters; \gps$\to UV_{uvw1}$\footnote{This is 
 approximately the UVW1 SWIFT filter bandpass. We chose this as it is
now the most commonly used NUV filter band-pass for transients and is reasonably
close to the restframe wavelength covered by the $g_{\rm P1}$ filter for $z=0.524$.},  \rps$\to u_{SDSS}$, \ips$\to g_{SDSS}$, \zps$\to r_{SDSS}$, \yps$\to i_{SDSS}$.
Details of all the photometry and \emph{K}-correction data for \ap\ can be found in Appendix\,\ref{app:A}.

\begin{figure*}
\includegraphics[angle=270,scale=0.6]{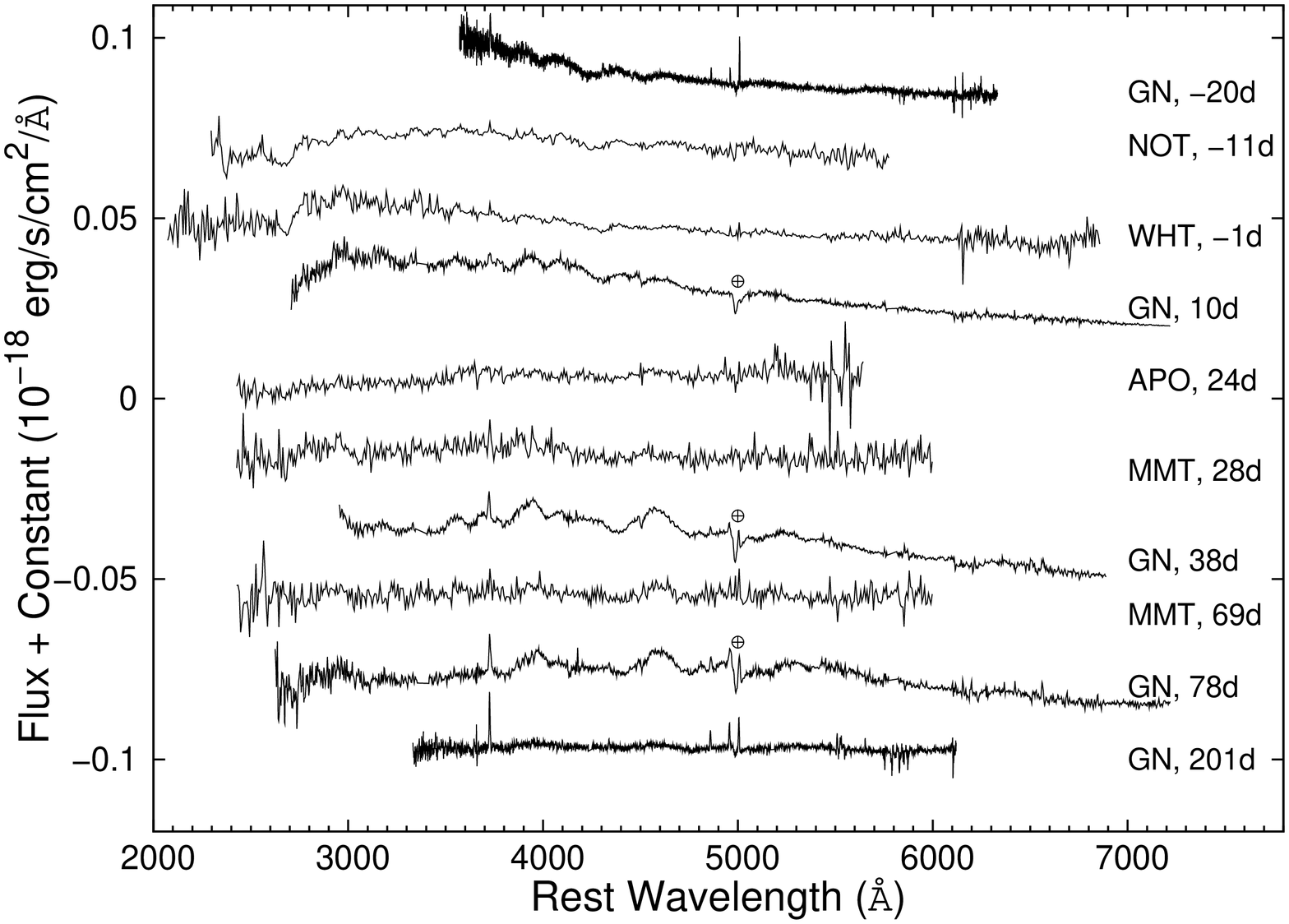}
\caption{The complete spectral series for \ap\ in the galaxy restframe.  Telluric features that have not been removed are marked with a `$\oplus$'.}
\label{fig:11apspec}
\end{figure*}

\begin{table*}
 \caption{Information on all of the spectra obtained for \ap.}
 \label{tab:11ap_spec}
 \begin{tabular}{@{}lccccccc}
  \hline
  \hline
Date & MJD & Phase (rest) & Telescope & Grating & Wavelength range (\AA) & Resolution (\AA) & P.I. \\
    \hline
24/01/2011 & 55585 & -20 days & GN + GMOS & R400 & $5440-9650$ & 5 & J. Tonry \\
06/02/2011 & 55598 & -11 days & NOT + ALFOSC & Gm14 & $3500-8800$ & 13 & R. Kotak \\
22/02/2011 & 55614 & -1 days & WHT + ISIS & R300B; R158R & $3160-10,500$ & 6 & S. Smartt \\
11/03/2011 & 55631 & +10 days & GN + GMOS & R150 & $4120-10,500$ & 12 & A. Pastorello \\
31/03/2011 & 55651 & +24 days & APO + DIS & B400; R300 & $3700-9000$ & 7 & M. Huber \\
06/04/2011 & 55657 & +28 days & MMT + HECTOSPEC & 270 & $3700-9150$ & 5 & E. Berger \\
22/04/2011 & 55673 & +38 days & GN + GMOS & R150 & $4500-10,500$ & 12 & A. Pastorello \\
08/06/2011 & 55720 & +69 days & MMT + HECTOSPEC & 270 & $3700-9150$ & 5 & E. Berger \\
22/06/2011 & 55734 & +78 days & GN + GMOS & R150 & $4000-10,500$ & 12 & D. A. Howell \\
27/12/2011 & 55922	& +201 days & GN + GMOS & R400 & $5000-9320$ & 5 & S. Smartt \\
\hline
 \end{tabular}
 \medskip
\end{table*}

\subsection{Comparisons with SLSNe}
\label{photcomp}

A photometric comparison of \ap\ with the SLSNe \dam, \bi, \xk\ and \gx\ \citep{12dam,07bia,07bib,11xk,10gx,quimb}  
is shown in Fig.\,\ref{fig:11apab}. 
To compare the respective absolute AB magnitude of each supernova we used the following:

\begin{equation}
M=m-5log(\frac{d_{L}}{10(pc)})+2.5log(1+z)
\label{eq:abmag}
\end{equation}

\noindent \citep{Hogg}, where \emph{m} is the apparent AB magnitude.  This corrects the measured magnitudes for cosmological expansion but is not a proper K-correction.  To make the comparison between different objects valid, appropriate filters were chosen so that the central rest wavelengths were as similar as possible (details of this are shown in Table \ref{tab:11ap_k}).  Note that although full K-correction data for \ap\ is given in this paper (see Appendix\,\ref{app:A}) and for \dam\ in \cite{12dam}, all the SLSNe data used here are treated with the same pseudo-correction given in Eq.\,\ref{eq:abmag} to ensure a reasonably consistent comparison.  A standard cosmology with $H_0=72$ kms$^{-1}$, $\Omega_M=0.27$ and $\Omega_\lambda=0.73$ is used throughout.

Two distinct groups of objects can clearly be seen in this plot; those transients with steeply-declining light curves post peak and those with long-lived light curves, falling by no more than $\sim$ 2 mag in 100 days.
\ap\ seems to belong to this latter group. The only SN to
previously fall definitively into this class was \bi. \cite{07bia} suggested that
the shallow gradient was powered by the radioactive decay of $^{56}$Ni
$\to$ $^{56}$Co $\to$ $^{56}$Fe synthesised as a result of a PISN. The
recent discovery of \dam\ \citep{12dam} and the data for \ap\ presented in this
paper allow us to shed new light on this class of objects.

Fig.\,\ref{fig:11apab} also shows a comparison of a \ap\ \gps-band
light curve with an \emph{i}-band light curve of the high redshift
($z=2.046$) SN2213-1745 \citep{cooke}, \emph{g} and \emph{u}-band
light curves of the low redshift ($z=0.107$) \dam\ \citep{12dam} and
PISNe model fits for a 100\msun\ and a 130\msun\ bare helium core
\citep{pisnmod}. The PISNe model light curves were calculated by
applying a synthetic SDSS \emph{u}-band filter ($\lambda_{c}=3550$\AA)
to the model PISN spectra from \cite{pisnmod} at all epochs using the
\emph{synphot} function within \textsc{iraf}. This gave model observed
light curves in magnitudes which could be offset to the same peak
absolute magnitudes as the SLSN data by subtracting an arbitrary
constant from each model. No scaling for time dilation (as the observed data were already corrected to rest frame) or colour
corrections were applied.
The data for \ap, \dam\ and SN2213-1745 all illustrate that their light curves may be similar.
It is possible that SN2213-1745 is akin to these lower redshift SLSN counterparts. The differences in the peak magnitudes could be due to the nature of
the pseudo-correction applied or evidence of intrinsic variation within this group of SLSNe.
However this conclusion relies on the assumption that the missing epochs for SN2213-1745 would follow the same trend as \ap\ and \dam.  We acknowledge that this is somewhat speculative. Given the high redshift of the \cite{cooke} PISN candidate, the central rest wavelengths compared are also quite different.  The \cite{cooke} data probe further into the UV than the data in this paper or in \cite{12dam} which must be taken into account regarding any statements on the physical nature of SN2212-1745.
In summary, the rise times of SN2213-1745, \ap\ and \dam\ appear similar and the peak magnitudes and decay times are also not dissimilar. It is possible, but not confirmed, that all three objects belong in the same class.

This highlights the uniqueness of the \ap\ dataset and the importance
of comparing data with similar central rest wavelengths, hence the
inclusion of the \emph{u}-band \dam\ points around peak. The observed
light curves of \ap\ (Fig.\,\ref{fig:11ap_lc}) show a marked
difference in the evolution of the bluer bands. The observed \gps-band
light curve, corresponding to an ultraviolet rest wavelength (see
Table\,\ref{tab:11ap_k} for the central rest wavelengths of all the
SLSNe mentioned and their corresponding observed filters), declines
much more quickly than the red filters which fall into the NUV and
optical regions  after redshift
corrections. The comparison seen in Fig.\,\ref{fig:11apab} shows that
\ap\ and \dam\ have a similar light curve evolution in the
$\sim$NUV/optical filters but the higher redshift of \ap\ allows us a
look into the UV that bridges the gap between the low redshift objects
of this SLSNe class (2007bi and \dam), and the high redshift examples
in \cite{cooke}. Unfortunately \emph{u}-band data for \dam\ was only
available for a short period around the peak epoch. \dam\ has the most
comprehensively observed rise time data of an object of this supposed
class but \cite{12dam} also find that PISN models are poor fits to the rise time
data for \dam\ with PISN models.

\section{Spectra}
\label{spectra}
\subsection{Dataset}

\begin{figure*}$
\begin{array}{cc}
\includegraphics[angle=270,scale=0.35]{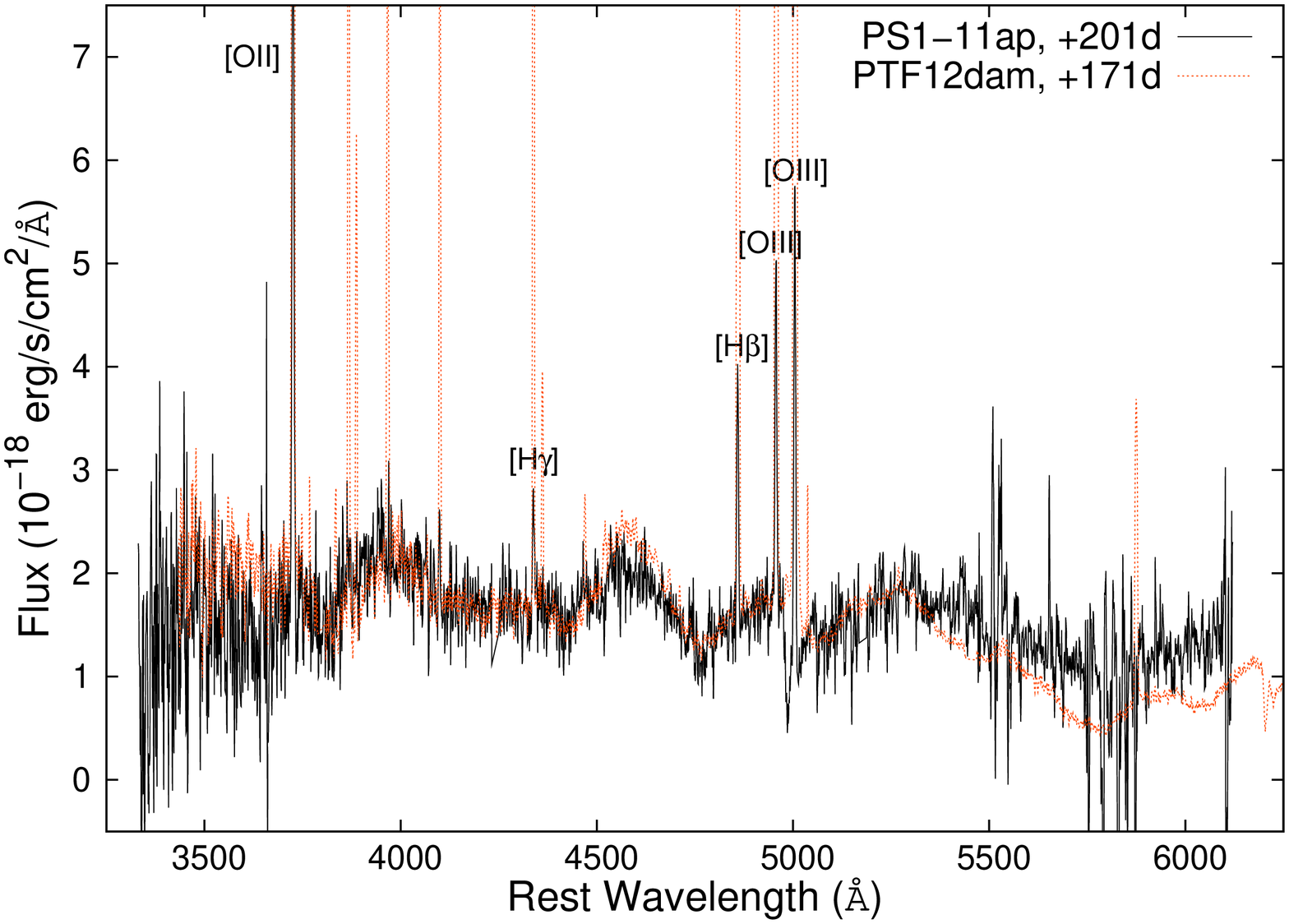} &
\includegraphics[angle=270,scale=0.35]{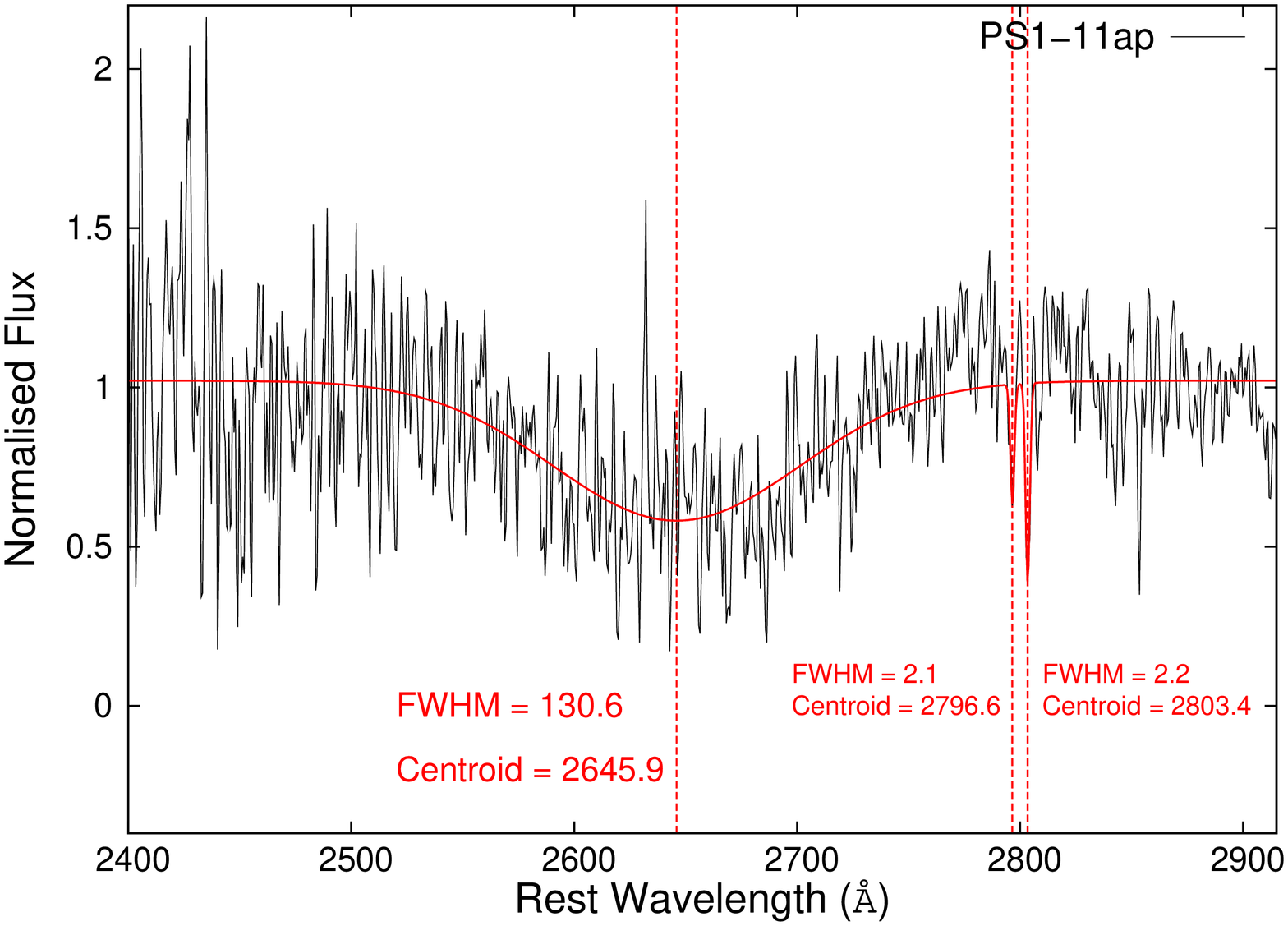} \\
\end{array}$
\caption{The left-hand panel shows the narrow emission lines in a late time GMOS-N spectrum for \ap\ used to infer the redshift of the host galaxy as $z=0.524$.  Another GMOS-N spectrum for a late epoch of \dam\ is overplotted to highlight the similarities between the objects even at this late stage.  The right-hand plot is a detail of the WHT \ap\ spectrum showing the rest wavelength of the two absorption features thought to be the Mg\,{\sc ii}  $\lambda\lambda$2796, 2803 doublet used to confirm the redshift value of the SN.  The red line overplotted here shows the three Gaussian fits used (any values listed on this plot are in \AA).}
 \label{fig:11apz}
\end{figure*}

Optical spectra of \ap\ were obtained with the Gemini North (GN)
telescope with GMOS-N\footnote{Program identification numbers: \emph{GN-2011A-Q-8}, \emph{GN-2011A-Q-16} and \emph{GN-2011B-Q-75}}, Nordic Optical Telescope (NOT) with the ALFOSC
spectrograph and the William Herschel Telescope (WHT) with ISIS.
Additional data were also provided from the Multiple Mirror Telescope
(MMT) and the Apache Point Observatory (APO).  In total, 11 epochs were
obtained including a late-time GMOS-N spectrum obtained on 27 December 2011 (during
the second season)  in an attempt to gather data on either late-time spectra of the SN and/or the host galaxy. During
the \ap\ season 1 at least one spectrum was obtained for each month
that the SN was visible. Details of the spectral series are given
in Table \ref{tab:11ap_spec}.

The WHT and NOT spectra were extracted using the \textsc{iraf}
\emph{apall} task after the usual de-bias and flat-fielding
procedures. The GN spectra were extracted using the custom built
\emph{gemini} \textsc{iraf} package. All epochs were wavelength
calibrated using spectra of CuAr lamps taken on the same respective
epoch as the science frames and flux calibrated using observations of spectroscopic standard stars taken on epochs as close to the science frames as possible. 
Both the \ap\ GMOS-N spectra obtained on the 11th March and on the 22nd April were
calibrated using a Feige34 spectrum taken on the 22nd April. The
spectrum obtained on the 22nd June was calibrated using an observation
of Feige34 from the same night and the spectrum from the 27th December
was calibrated using a spectrum of Feige34 obtained on the 24th
December. All the spectra for \ap\ can be seen in
Fig.\,\ref{fig:11apspec}.

A redshift value of $z=0.524$ was initally derived for \ap\ from our
first GMOS-N spectrum (taken in January 2011) from the narrow emission
lines from the SN host galaxy. These are seen throughout
the spectral series, most apparently in the late time GMOS-N spectrum
as seen in Fig. \ref{fig:11apz}. The use of the R400 grating for this
early exposure however meant that a restframe wavelength range of just
over 2500$\AA$ was obtained. Two spectra were obtained in February
using the ALFOSC spectrograph and the ISIS spectrograph which both
extend bluewards of $\sim3500\AA$. This redshift means that the WHT restframe spectrum
covers the
Mg\,{\sc ii} $\lambda\lambda$2796, 2803 doublet. This was initially used to
determine the redshift of multiple SLSNe by \cite{quimb} and \cite{chom} and subsequently 
\cite{berg} illustrated its use in providing diagnostics of the ISM in the
apparently unusual host galaxies of SLSNe. 
The redshift of  $z=0.524$  was confirmed from simultaneous Gaussian fits to both  
components of the
 Mg\,{\sc ii} $\lambda\lambda$2796, 2803 doublet, indicating that the 
absorption is very likely to be from the ISM of the host galaxy of \ap. By
adopting the Mg\,{\sc ii} centroids as rest frame the wider, bluer profile (assumed to be
due to the supernova)  could be fit with a simple Gaussian 
absorption profile (seen in Fig. \ref{fig:11apz}). This provided an expansion
velocity of $\sim16, 500$\,kms$^{-1}$ which is 
similar to SLSNe which have previously been studied :  velocities
ranging from $10, 000-20,000$\,kms$^{-1}$ were found in \cite{quimb,chom} and \cite{11xk}.

Three more GMOS-N spectra were taken in March, April and June this
time utilising the R150 grating to increase the wavelength range of
the exposures at the expense of some resolution. The APO and MMT data
also offer some completeness around these epochs, although the low signal-to-noise 
limits their diagnostic usefulness. Finally, one more
spectrum was obtained during the season 2 of \ap\ using the R400
grating to provide a resolution better matched to measuring the narrow
galaxy emission lines ($\sim300$\,kms$^{-1}$).  This was taken at a rest frame epoch of 201 days.

\begin{figure*}
\includegraphics[angle=270,scale=0.55]{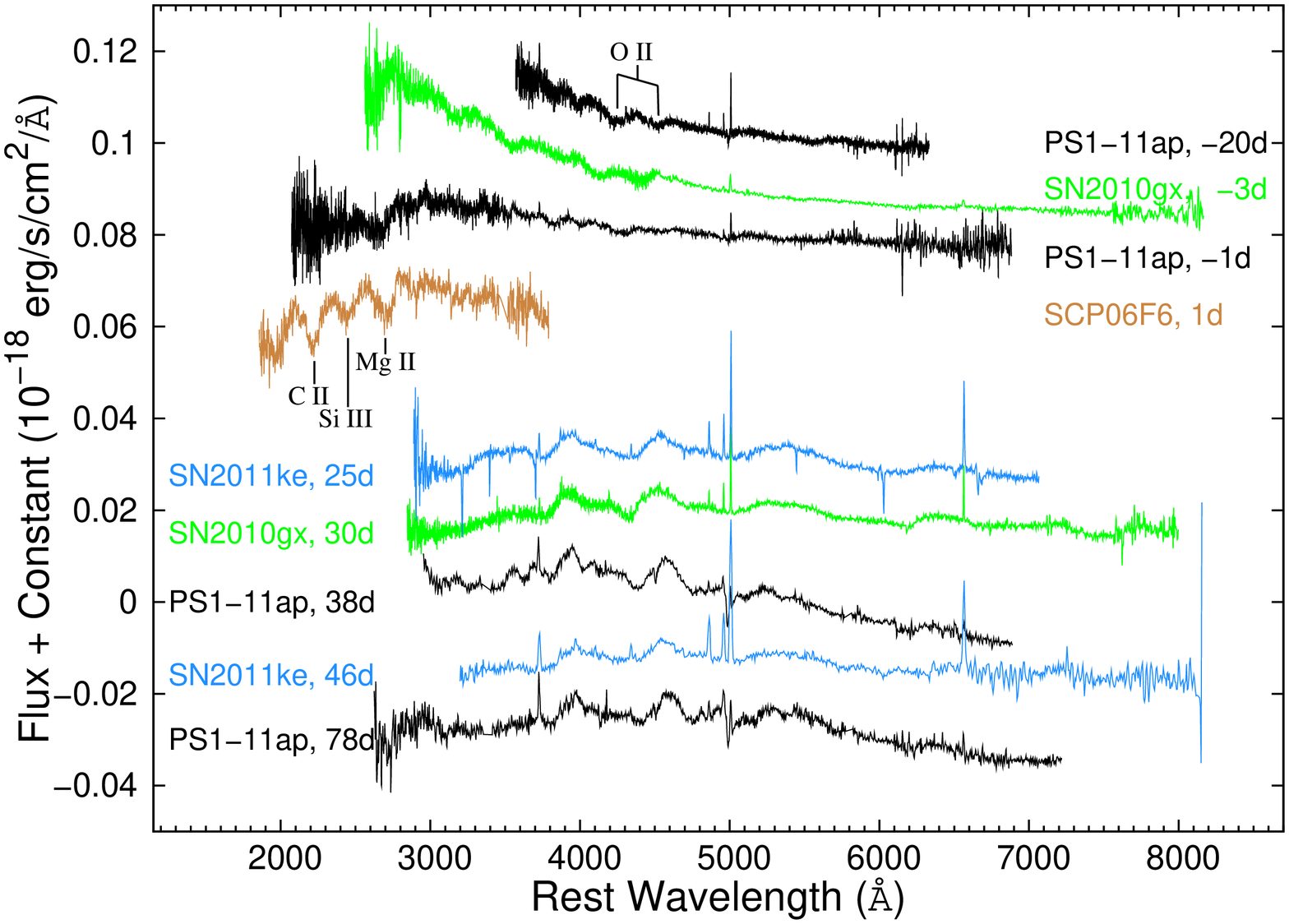}
\caption{A spectral comparison of \ap\ with the SLSN Ic \gx, SCP06F6 and \xk.}
 \label{fig:11apspeccomp2}
\end{figure*}

\begin{figure*}
\includegraphics[angle=270,scale=0.55]{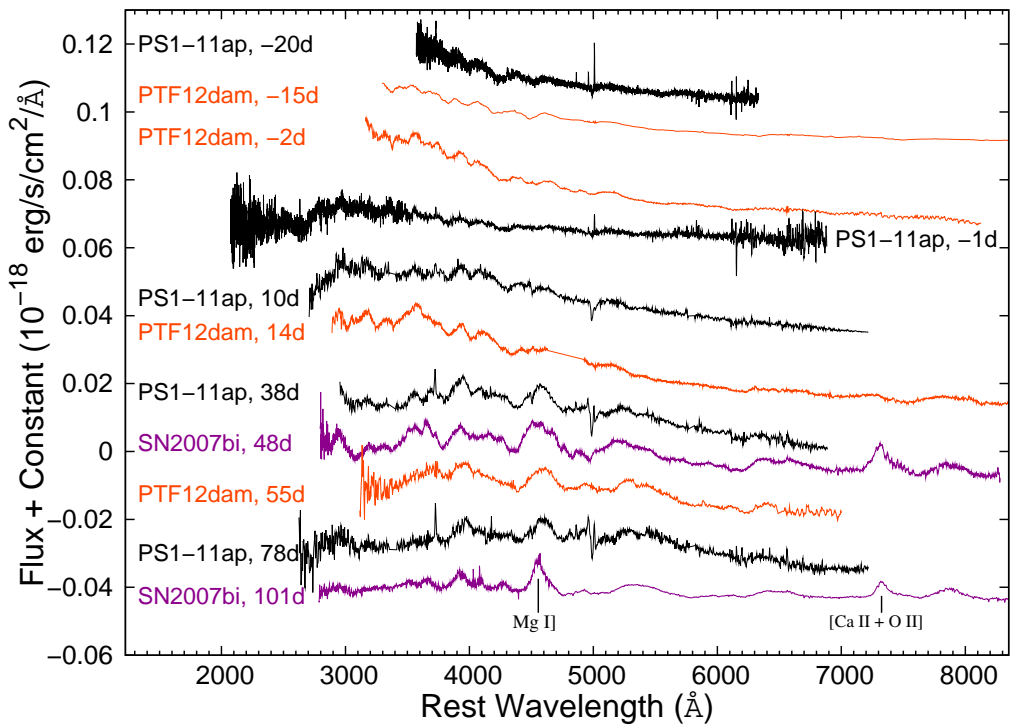}
\caption{A spectral comparisons of \ap\ with \dam\ and \bi.  Despite having spectral similarities with both \bi-like objects and SLSN Ic we group \ap\ with the former due to its photometric evolution.}
 \label{fig:11apspeccomp}
\end{figure*}

\subsection{Spectral comparisons}

\cite{quimb} and \cite{chom} identified three broad absorption features in SLSNe-Ic at
$\sim2330\AA$ (C\,{\sc ii}), $2543\AA$ (Si\,{\sc iii}) and $2800\AA$
(Mg\,{\sc ii}) as exemplified in the SCP06F6 spectrum \citep{quimb,scp06f6} in
the comparison plot in Fig.\,\ref{fig:11apspeccomp2}. Here we see how
the \ap\ data compares with data for the SLSNe-Ic \gx, SCP06F6 and \xk\
\citep{10gx,scp06f6,11xk}.  Between peak and 80 days, the 
\ap\ spectra are quite similar to these SLSNe-Ic. 
Two epochs of the \ap\ spectral series push bluewards
of $\sim2500\AA$ and  broad Mg\,{\sc ii} is seen in absorption in  \ap\ (highlighted in the
right hand plot in Fig. \ref{fig:11apz}). Unfortunately neither the
\bi\ nor the \dam\ spectra probe into the NUV. Another
common feature that \cite{quimb} identified with the SLSN-Ic class is a
broad `W'-shaped feature at $\sim4300\AA$ (O\,{\sc ii}). This feature
is also seen in the spectra of \ap\ and \dam. The photometric
evolution of \ap\ and objects of this class are very different,
however, as previously discussed in Section\,\ref{photcomp}.

We do not show a spectral comparison of \ap\ with objects from the SLSN-II class,
such as SN2006gy \citep{06gya,06gyb}, as none of the
\ap\ spectra show any evidence of the broad H and He emission shown by
SNe of this type.

The intermediate redshift of \ap\ makes a direct spectral comparison
between it and \bi\ and \dam\ (both at redshifts of z$\sim0.1$)
difficult. Spectra for \ap\ are nevertheless compared with
spectra for \dam\ at all corresponding epochs and with spectra for
\bi\ at two later epochs \citep{12dam,07bib}.  Both the -20 day \ap\
spectrum and the -15 day \dam\ spectrum (see
Fig.\,\ref{fig:11apspeccomp}) share a common blue colour with shallow, broad absorption features. 
The later spectra, especially around $\sim50$
days, match very closely for all three objects. This similarity,
combined with the photometric similarities in the slowly declining and
very luminous light curves, provides convincing evidence that these
objects are of the same physical class. \cite{07bib} note that the
spectra for \bi\ are like that of a slowly evolving Type Ic SN but with
an extremely extended photospheric phase that only begins its
transition into the nebular phase at $350-400$ days post-discovery. Of
interest is the appearance of nebular features, seen in the clear
emission lines at $\sim4500\AA$ (\Mgi]) and $\sim7320\AA$ (thought to
be associated with the forbidden \Caii\ $\lambda\lambda$7291, 7324
doublet and a forbidden emission feature of \Oii\ at $7322\AA$) in
both the 48 day and the 101 day \bi\ spectra shown in
Fig.\,\ref{fig:11apspeccomp}, despite the clear presence of a retained
continuum. When corrected for redshift, the wavelength coverage of the
optical spectra for \ap\ only covers the \Mgi] feature but it is also
seen in the post peak epochs although not as strongly as in the case
of \bi.

A more detailed discussion on the nature of \ap\ is given in Section\,\ref{disc}.

\subsection{The Host Galaxy of PS1-11ap}
\label{host}

\begin{table}
\begin{center}
 \caption{Main properties of the host galaxy of PS1-11ap. Foreground
   extinction was applied to determine of the
 absolute AB magnitudes at effective restframe noted. }
 \label{tab:hostmag}
 \begin{tabular}{@{}lc}
    \hline    \hline
PS1 Name & PSO J162.1155+57.1526 \\
\gps (mag)& 24.10 (0.32)                                          \\ 
\rps (mag)&  23.16 (0.17)                                         \\ 
\ips (mag)&  22.88 (0.15)                                          \\ 
\zps (mag)& 22.90 (0.24)                                          \\ 
Internal extinction $A_{V}$  (mag) &              $\sim0$               \\ 
$M_{\rm 316}$ (mag)&         -17.79           (0.32)                    \\ 
$M_{\rm 405}$ (mag)&                     -18.72          (0.17)          \\ 
$M_{\rm 493}$ (mag)&                                 -19.00       (0.15)\\ 
$M_{\rm 568}$ (mag)&                                      -18.97  (0.24)\\ 
Physical diameter (kpc) &           3.6                              \\ 
Luminosity [O\,{\sc ii}] (erg s$^{-1}$) &   $7.20\times10^{40}$                                     \\ 
SFR (M$_{\odot}$\,yr$^{-1}$) &      $0.47\pm0.12$                                   \\ 
Stellar Mass (M$_{\odot}$) &     $1.5^{+1.40}_{-0.65}\times10^{9}$                                   \\ 
sSFR (Gyr$^{-1}$)         &         $0.31^{+0.24}_{-0.15}$                       \\ 
 \hline
 \end{tabular}
\end{center}
\end{table}

The host galaxy of \ap, \emph{PSO J162.1155+57.1526}, was detected in \gps, \rps, \ips\ and \zps-band \PS\ images made from a co-addition of
nightly images before the first discovery epoch of \ap.
The host was not significantly detected in \yps-band stacks.
The RA and Dec used for the position of the host galaxy were the same as that of \ap, given in Section\,\ref{phot}, as the SN was coincident with the galaxy centroid.

The \ips-band image (total exposure time of 3840s) shows that there is a
nearby fainter object at approximately 7 pixels from the centroid of
the host galaxy of \ap. As the host was extended and not well
described by a PSF, we chose to carry out aperture photometry
with a 7 pixel aperture and applied an aperture correction using 8 isolated
stars in the \PS\ image (within 2 arcmin of the position of \ap). 

\begin{figure}
\includegraphics[angle=270,scale=0.35]{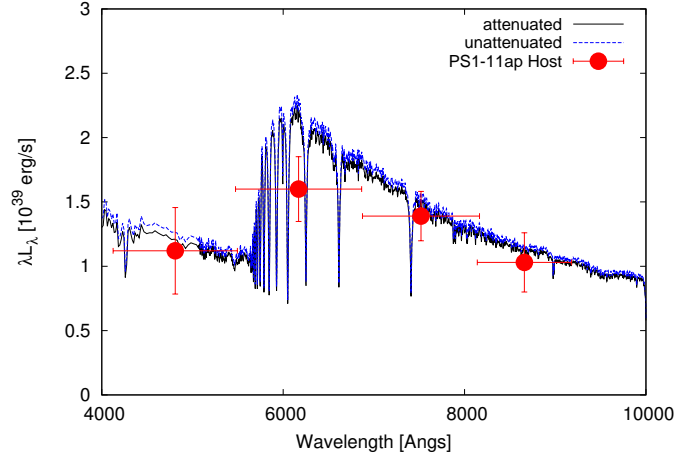}
\caption{Photometry of the host galaxy (red points) with the best-fit ($\chi^{2}$=0.011) model galaxy SED (black line)
from MAGPHYS (da Cunha et al. 2008), with and without an attenuation by internal dust.  This best fit model has a 
stellar mass of $7.4\times10^{8}$\msun, but the more realistic range of masses is given in Table 3.}
 \label{fig:magphys}
\end{figure}

We applied a correction for Milky Way extinction \citep{2011ApJ...737..103S} and  estimated the contribution of the nebular
emission lines (discussed below) to the broad band photometry. 
This was required in order to determine the galaxy mass from stellar
population model fitting to the observed galaxy colours. The 
correction was minimal and we found that the measured emission line 
strengths would only imply a change of 0.007 mag in \rps\ and 0.01 mag in \ips.   
The absolute magnitudes are listed in Table\,\ref{tab:hostmag} at the
effective restframe wavelengths of the \PS\ filters. A full
K-correction was not calculated for these, they are simply estimated
from Eq.\,\ref{eq:abmag}.
We used the MAGPHYS stellar population model code of \cite{2008MNRAS.388.1595D}
with our redshift of  $z=0.524$ with the goal of determining the stellar mass. 
The code returns a spectral energy distribution (SED) and best 
fit model, which is simply defined 
as the model with the lowest $\chi^{2}$ while allowing all the input parameters to 
vary. This lowest $\chi^{2}$ solution is plotted in Fig.\,\ref{fig:magphys} with the \ap\ host galaxy \emph{gri}\zps\ photometry where the x-axis error bars represent the \PS\ filter bandpasses.
The $\chi^{2}$ value is low ($\chi^{2}=0.011$) and not reflective of the range of acceptable
fits for stellar mass. MAGPHYS gives a more realistic range of
values for stellar mass by calculating the probability density function
over a range of values and determining the median and 
the confidence interval corresponding to the 16th$-84$th percentile range. 
This is equivalent to the 1$\sigma$ range, if the distribution is 
roughly Gaussian (see da Cunha et al. 2008 for details). As recommended
by \cite{2008MNRAS.388.1595D} we take the best estimate of stellar mass to 
be this median value, $1.5\times10^{9}$M$_{\odot}$, and the 1$\sigma$
range to be $0.85-2.9\times10^{9}$M$_{\odot}$.  
The best fit derived in Fig.\,\ref{fig:magphys} has a mass which is on the edge of the 
1$\sigma$ range,  $0.74\times10^{9}$\msun, as this fit comes from allowing all parameters to vary
and the range is from marginalising over stellar mass only. This illustrates
that only the range should be considered as a reliable estimate, not a particular best fit.

\begin{table}
\begin{center}
 \caption {Observed emission line strengths of the host of PS1-11ap
   taken by Gemini on 2011/12/27. The line strengths are as   measured
from the observed spectrum with no correction for redshift, extinction or distance
yet. The last column is the equivalent width, in angstroms.}
 \label{tab:hostlines}
\begin{tabular}[t]{lllllll}
\hline
\hline
Line & Flux $\pm$ Error & EW \\
 & (erg s$^{-1}$ cm$^{-2}$) & ($\AA$) \\
\hline
$[$OII$]$ $\lambda$3727 & $(6.90\pm0.02)\times10^{-17}$ & 39.82 \\
H$\delta$ $\lambda$4102 & $(5.47\pm0.14)\times10^{-18}$ & 3.16 \\
H$\gamma$ $\lambda$4340 & $(1.02\pm0.05)\times10^{-17}$ & 6.31 \\
H$\beta$ $\lambda$4861 & $(2.00\pm0.03)\times10^{-17}$ & 12.79 \\
$[$OIII$]$ $\lambda$4959 & $(2.75\pm0.02)\times10^{-17}$ & 15.20 \\
$[$OIII$]$ $\lambda$5007 & $(4.50\pm0.02)\times10^{-17}$ & 55.81 \\
\hline
\end{tabular}
\end{center}
\end{table}

\begin{figure*}$
\begin{array}{ccc}
\includegraphics[angle=270,scale=0.3]{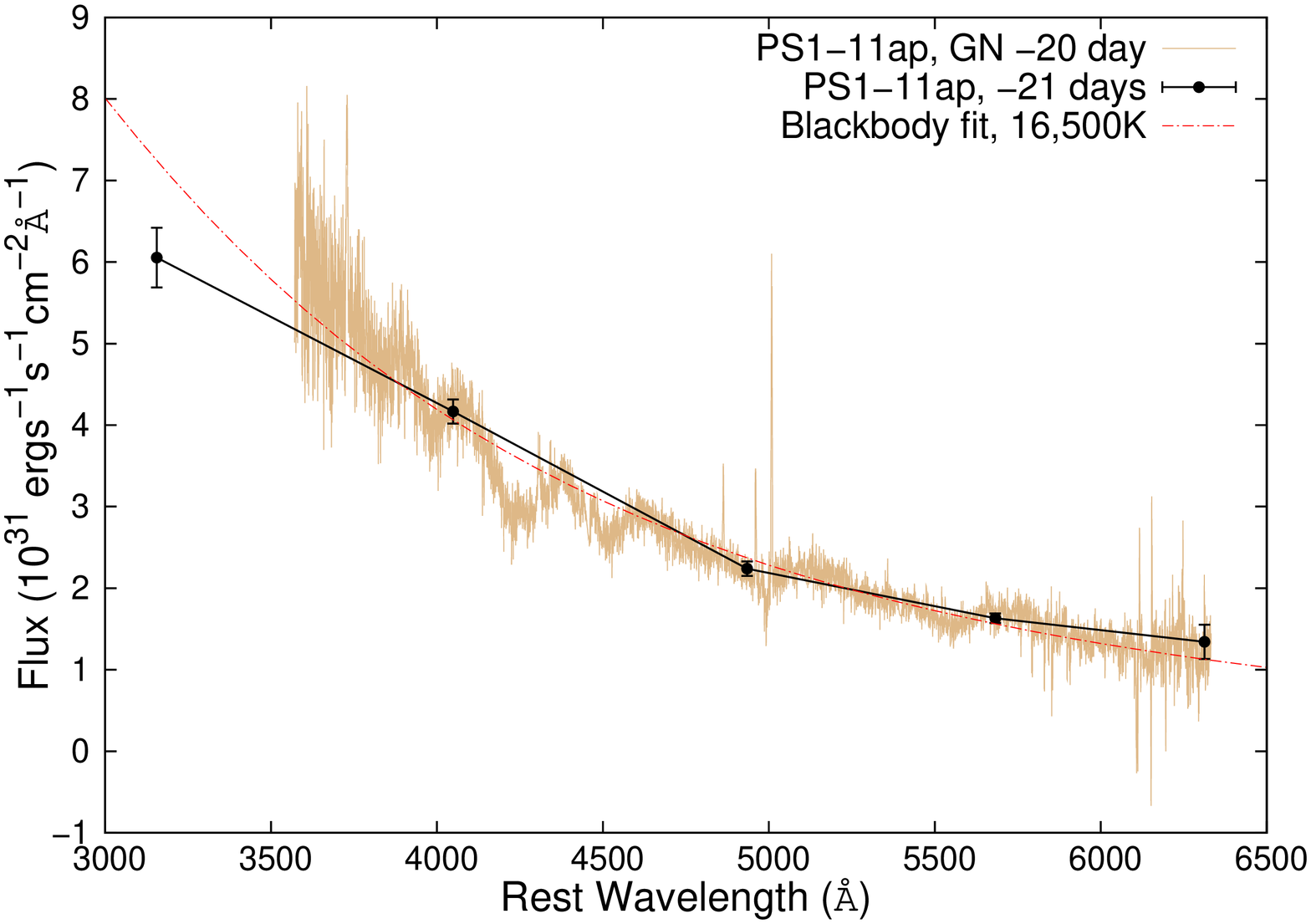} &
\includegraphics[angle=270,scale=0.3]{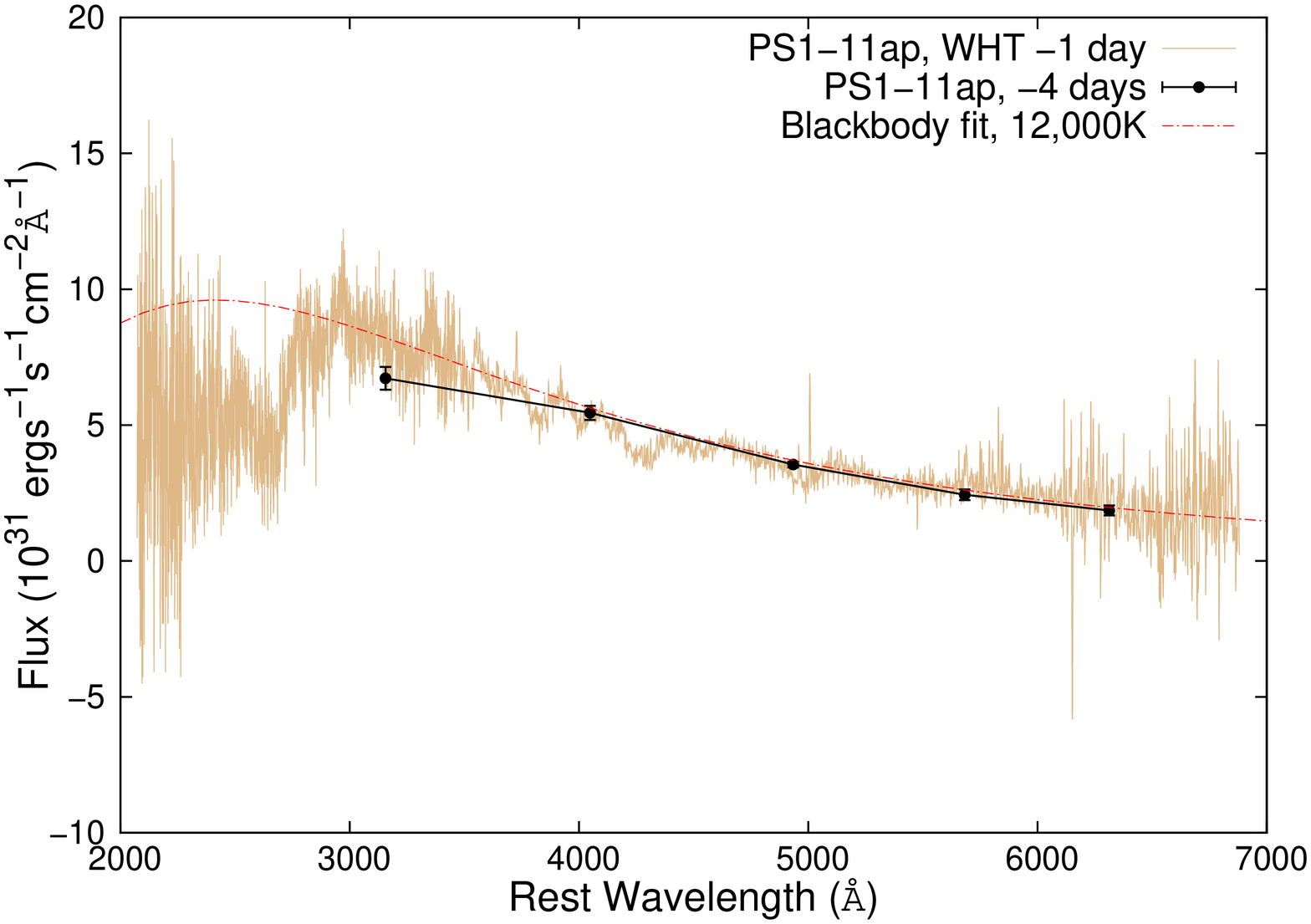} \\
\includegraphics[angle=270,scale=0.3]{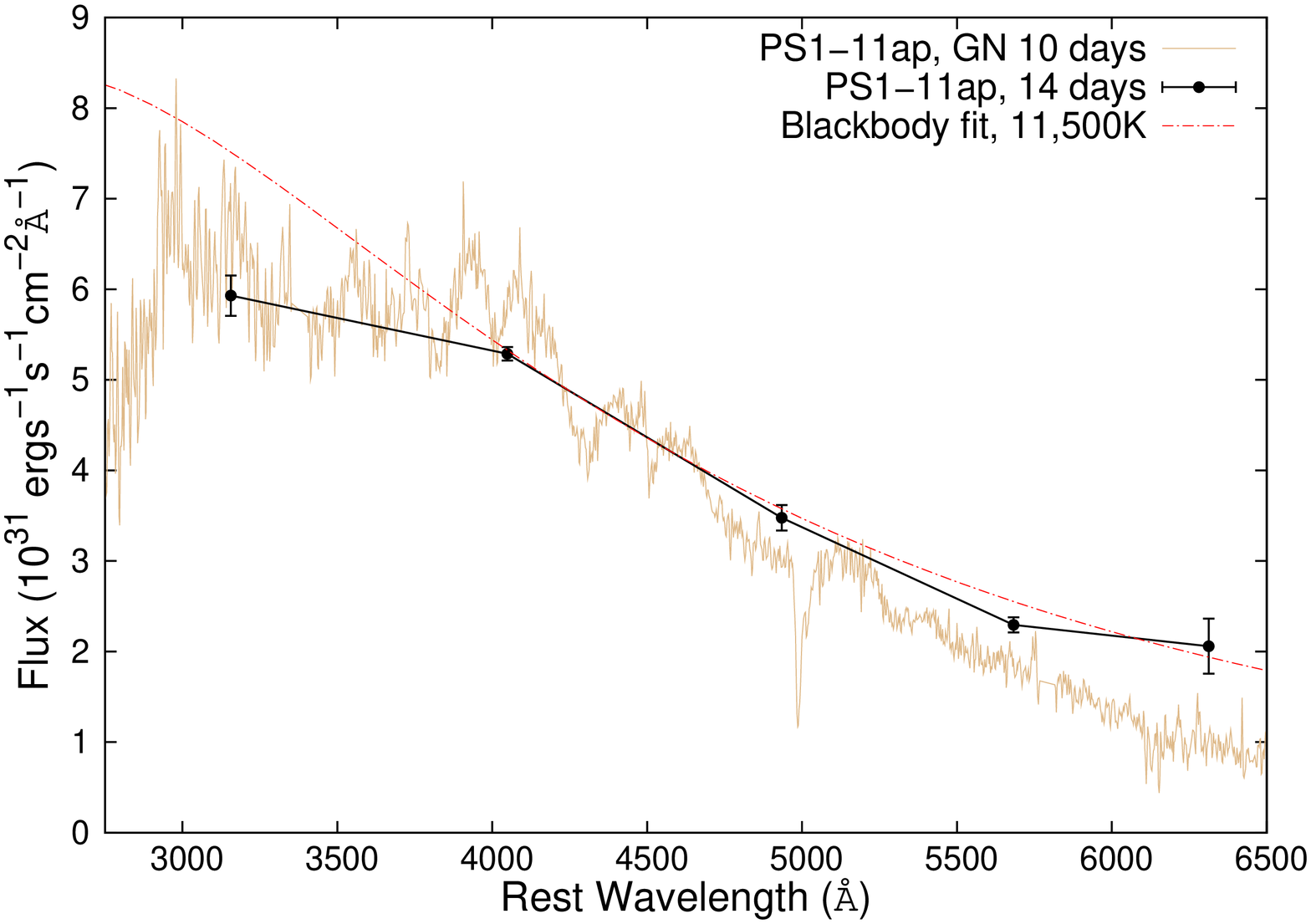} &
\includegraphics[angle=270,scale=0.3]{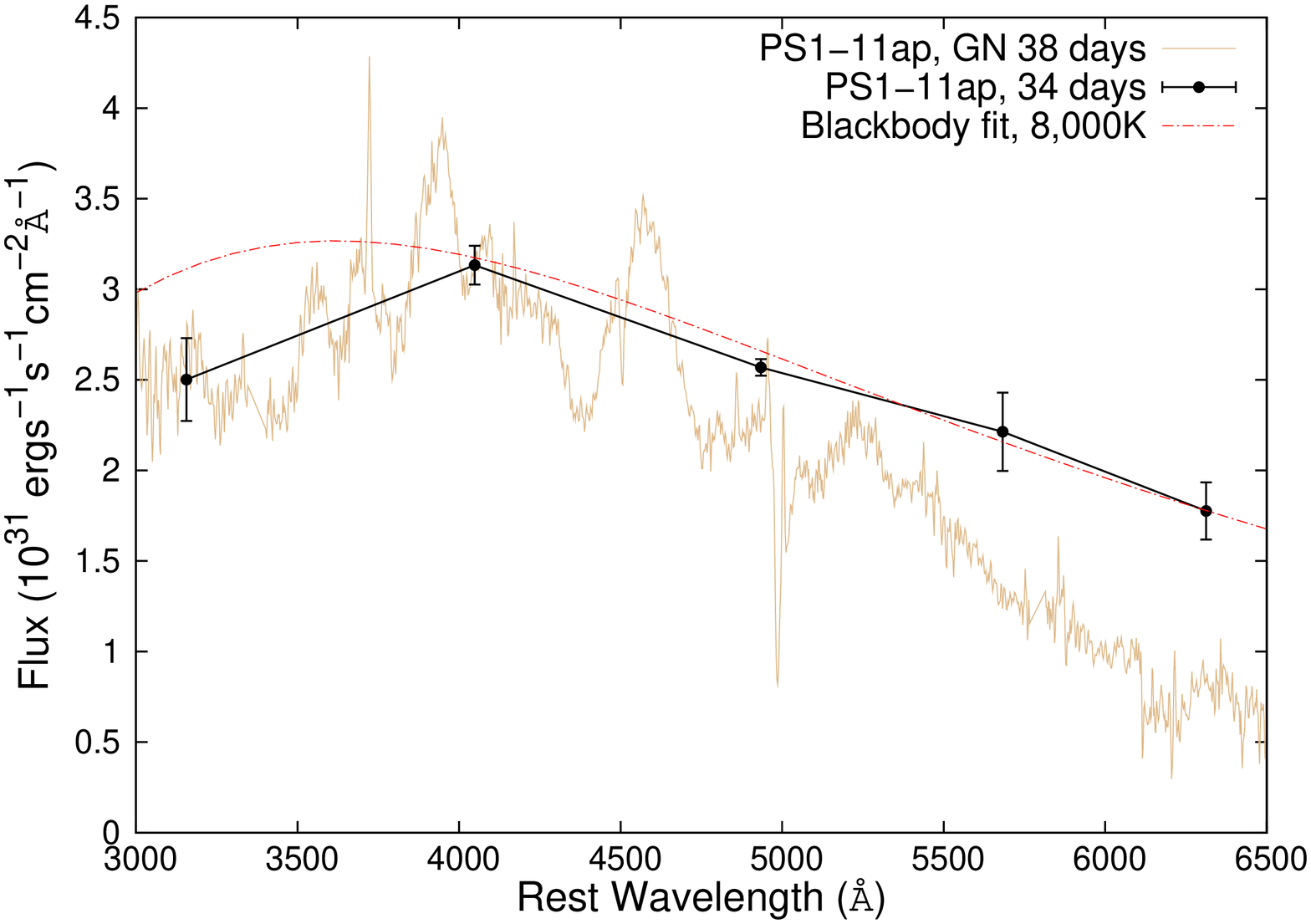} \\
\includegraphics[angle=270,scale=0.3]{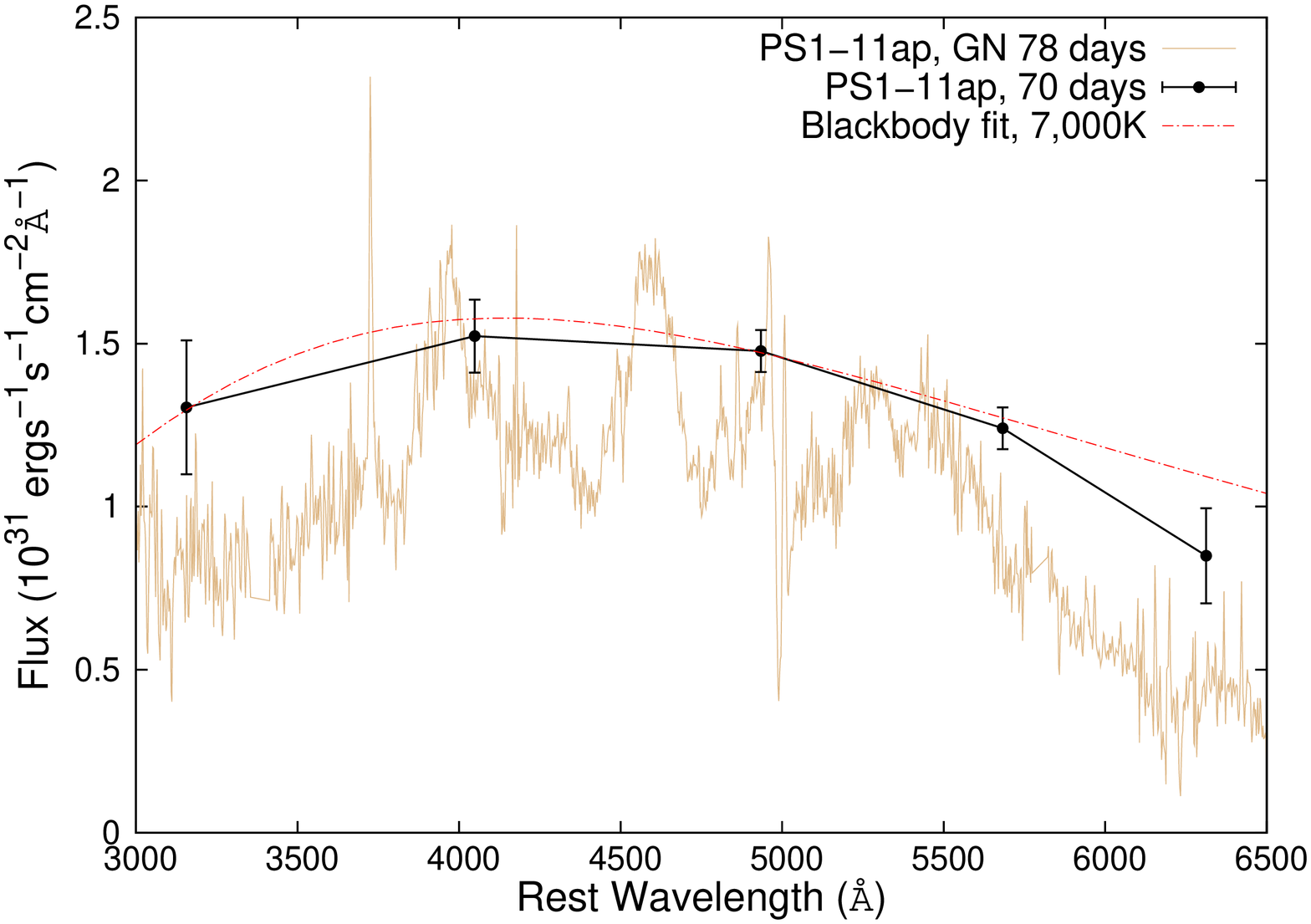} &
\includegraphics[angle=270,scale=0.3]{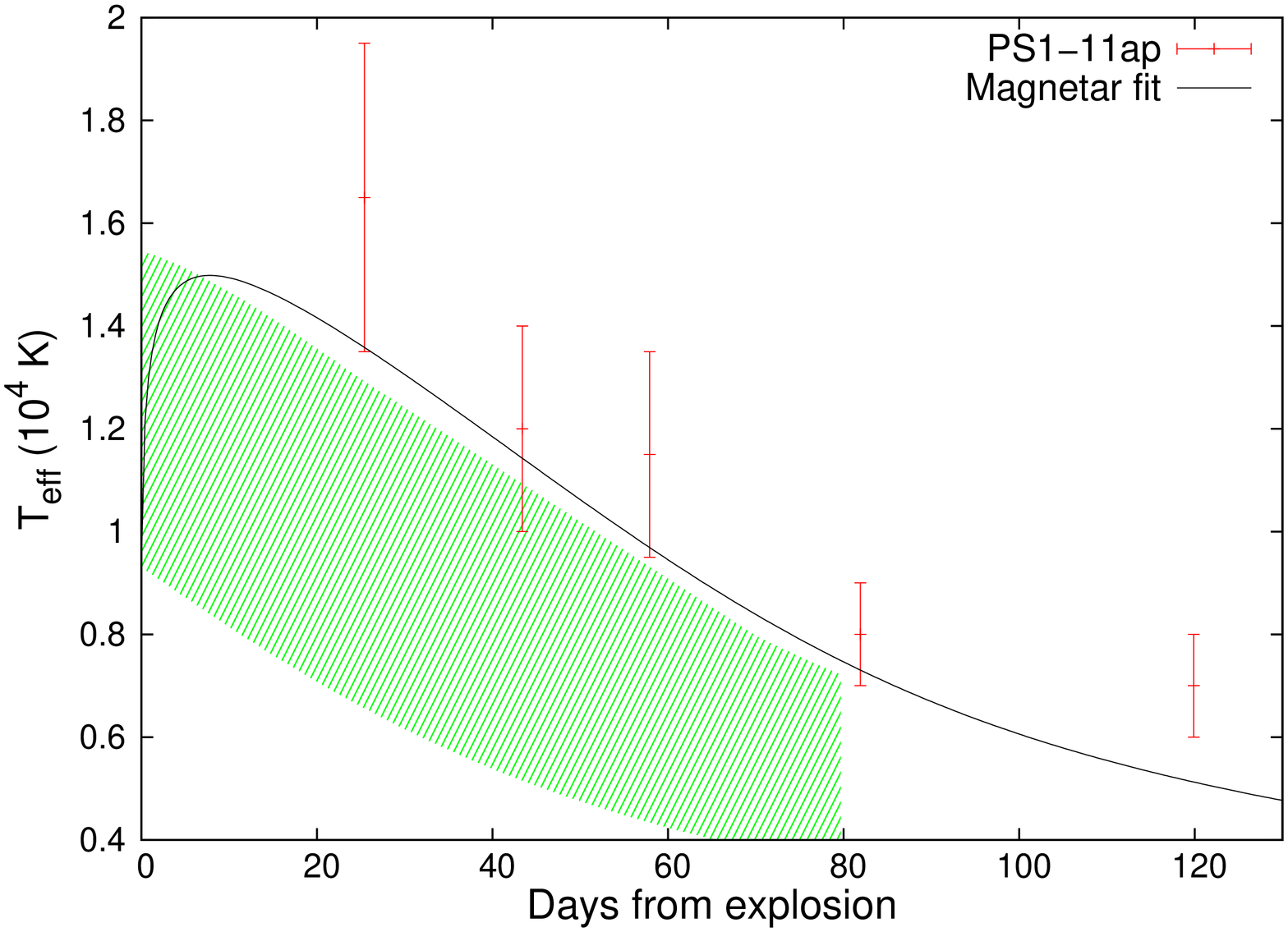} \\
\end{array}$
\caption{Blackbody fits to the \ap\ photometry with similar epoch spectra overplotted for consistency.  The lower right figure shows how the fitted temperatures compare with the magnetar model fit from the lower right hand plot in Fig.\,\ref{fig:11apfits}, where the green shaded region covers the data of the four SLSN-Ic presented in Inserra et al (2013).}
 \label{fig:11apbb}
\end{figure*}

The final spectrum taken on 27 December 2011 with Gemini + GMOS still
has clear signs of the broad lines from PS1-11ap (see
Fig.\,\ref{fig:11apz}).  At this epoch of 201 days (restframe) the
detected continuum is a combination of the host galaxy continuum
(\rps$=23.16\pm0.17$ from the pre-2011 data) and supernova flux
(\rps$=23.68\pm0.17$) measured in the reference subtracted images. Emission
lines from the host galaxy are detected but at this redshift and
without an infrared spectrum we are limited to the strong nebular
lines visible between restframe wavelengths of $3700-5100$\AA, as listed in Table\,\ref{tab:hostlines}.
We fitted high order polynomials to the continuum and subtracted off
this flux before measuring the emission line fluxes. 
Line fluxes were then measured using the QUB custom built $procspec$ environment within IDL
and single Gaussian profiles were fit to each of the six identified features
listed in Table\,\ref{tab:hostlines}.  Although the [O\,{\sc ii}]
feature at 3727\AA\ is a doublet blend, the two components were not
resolved and a single Gaussian, broader than the other lines was
employed.  These are the observed line fluxes
before any extinction or redshift corrections were applied. 
There is no detection of the
electron temperature sensitive line [O\,{\sc iii}] 4363\AA, as used by
\cite{2013ApJ...763L..28C} and \cite{lun} to determine abundances in the host
galaxies of SN2010gx and PS1-10bzj. 
Hence we used the strong line $R_{23}$
method to estimate the oxygen abundance in the host.

To determine the intrinsic line emission strengths we first  applied a correction for the Milky Way foreground extinction of $A(V)=0.02$
and $R_{V}=3.1$ \citep{2011ApJ...737..103S}.  As we lacked H$\alpha$ in our
optical spectra due to the redshift,  we use the ratio of 
H$\gamma$/H$\beta$ to determine a value for internal extinction. 
The intrinsic line ratio assuming case B recombination for
H$\gamma$/H$\beta=0.47$ and for H$\delta$/H$\beta=0.26$ \citep{Oster89},
which implies an internal dust extinction of $A(V) \sim0$.
The resultant spectrum was then shifted to rest 
wavelength for line measurements. 

The reddening-corrected flux of the [O\,{\sc ii}] $\lambda3727$ line is $7.07\times10^{-17}$ (erg s$^{-1}$ cm$^{-2}$) and so
we determined the star formation rate (SFR) of the host to be $0.47\pm0.12\,\msun$\,year$^{-1}$ from the relation of \cite{2004AJ....127.2002K}.
Our restframe spectral coverage for \ap\ does not cover the wavelength of H$\alpha$ but as we know the intrinsic line ratio of H$\alpha$/H$\beta=2.86$ and have a measurement of H$\beta$, we can imply that the flux of H$\alpha=5.86\times10^{-17}$ erg s$^{-1}$ cm$^{-2}$.
From the calibration of \cite{1998ARA&A..36..189K}, SFR ($M_{\odot}$\,year$^{-1}$)$=7.9\times10^{-42}$ L(H$\alpha$) where the latter is in units of ergs\,s$^{-1}$, we calculate a SFR $\sim0.47\pm0.02\,\msun$\,year$^{-1}$, which shows a good agreement with the {\rm [O\,{\sc ii}]} result.

We determined the commonly used line ratio $R_{23}$ ([O\,{\sc iii}] $\lambda\lambda5007,4959$ + [O\,{\sc ii}] $\lambda3727$/H$\beta$)
finding $\log(R_{23})=0.86$.
This is close to the 
turnover region between lower and upper metallicity branches of the 
$R_{23}$ calibration plot and reduces our ability to constrain the
metallicity particularly well.  Using the \cite{1991ApJ...380..140M}
$R_{23}$  calibration, we determine  
$12+\log{\rm O/H}=8.5\pm0.3$\,dex for the upper branch, and 
$12+\log{\rm O/H}=8.1\pm0.5$\,dex  for the lower. The large errors reflect
the uncertainties from the line strength measurements.  
Typically there is also a systematic + 0.3\,dex offset between the oxygen abundance
determined with the $R_{23}$ ratio and the McGaugh (1991) calibration
and abundances determined on an electron temperature scale \citep{2011ApJ...729...56B}.
The $R_{23}$ calibration is uncertain and without a detection of an auroral line for electron temperature measurement (such as [O\,{\sc iii}] $\lambda4363$) we cannot determine the metallicity accurately.
It is therefore possible, but not definitive, that the host is of low metallicity similar to that 
determined by \cite{2013ApJ...763L..28C} and \cite{lun} for the hosts of SN2010gx and
PS1-10bzj respectively.

An additional probe of the host environment is ISM absorption of the
Mg\,{\sc ii} $\lambda\lambda$2796, 2803 doublet. This was a
key diagnostic used by \cite{quimb} and \cite{chom} to determine the redshifts of
early SLSNe discoveries and \cite{berg} illustrated the possible
usage of these lines as diagnostics of the higher redshift Universe.
Fig.\,\ref{fig:11apz} shows the Mg\,{\sc ii} absorption detected in
the WHT spectrum of PS1-11ap from the 22nd February, 2011.  The rest frame equivalent widths of the two lines are
$W_r(\lambda 2796)=0.8\AA$ and $W_r(\lambda 2803)=1.4\AA$ 
which are similar, within the errors, to those measured by 
\cite{berg}  for PS1-12bam. 

\subsection{Blackbody fits}

\begin{table}
\begin{center}
 \caption{Temperatures from blackbody fitting of \ap.}
 \label{tab:11ap_bb}
 \begin{tabular}{@{}lc}
  \hline
  \hline
Phase (rest) & $T_{bb}$ \\
    \hline
-20.5 days & 16500 $\pm$ 3000 K \\ 
-2.5 days & 12000 $\pm$ 2000 K \\
12 days & 11500 $\pm$ 2000 K \\
36 days & 8000 $\pm$ 1000 K \\
74 days & 7000 $\pm$ 1000 K \\
\hline
 \end{tabular}
\end{center}
\end{table}

Blackbody curves of various temperatures were fitted to the \ap\ photometry and spectra at 5 different epochs (listed in Table \ref{tab:11ap_bb}).  The effective wavelength of each of the \PS\ \grizy\ filters were taken from \cite{PS1phot} so that photometric data could be overplotted with an appropriate spectrum of similar phase.   Model blackbody curves for various temperatures could then be manually fitted where the error range was derived simply by choosing the range of fits which still satisfied the data within 3 times the photometric errors.
For example, a maximum and minimum temperature was selected for each epoch by trying to fit a range of blackbody curves to the photometric data until the fit passed through 3 times the error bars of less than 4 of the 5 effective wavelengths.  The estimated temperature for that epoch was then given as the midpoint of the temperatures represented by these model curves with an error of the difference between them divided by 2. Effectively, the quoted errors for the blackbody fits are to approximately 3$\sigma$.
It can be seen in Fig. \ref{fig:11apbb} that the model curves fit the \rps, \ips, \zps\ and \yps-band points well but that the \gps-band consistently falls short.  This may be due to line blanketing from heavier elements at this wavelength range, supported by the broad absorption features seen in the spectra here.

A correction of $E(B-V) = 0.006$ mag for Milky Way extinction in the direction of \ap\ from the NASA/IPAC IRSA dust maps\footnote{http://irsa.ipac.caltech.edu/applications/DUST/} \citep{2011ApJ...737..103S} was applied to the photometry in each filter and each spectrum was corrected using the \emph{dered} function in the \textsc{iraf} \emph{onedspec} package.  No correction was applied regarding the host galaxy of the supernova as the value of $A(V)\sim0$ derived in Section\,\ref{host} suggests a neglible average internal dust extinction.  Also, due to high redshift of \ap, the exact position of the SN within its host is not known and so any value would have to be averaged across the whole galaxy.  
In summary our temperatures derived in Table \ref{tab:11ap_bb} may suffer from a systematic uncertainty due to internal extinction which would only increase the intrinsic temperatures.

The lower right plot in Fig.\,\ref{fig:11apbb} shows how the evolution
of the photospheric effective temperature (T$_{eff}$) of the magnetar
model that best fits our bolometric light curve (see
Section\,\ref{mods}) matches our estimated T$_{eff}$ from the blackbody fitting of the obtained photometry and spectroscopy for \ap. As
found in \cite{12dam} the model seems to underestimate the derived
temperatures by about $10\%$, but still lies within the $1\sigma$ error
range of 4 of the 5 derived data points. The evolution of the model
matches the measured continuum temperature of \ap\  reasonably well, 
given the simplicity of the method. 
\cite{11xk} applied a similar method to a set of low redshift SLSNe-Ic 
and found quite similar trends for their temperature evolution. 
Fig.\,\ref{fig:11apbb}  shows the region occupied by 
 PTF10hgi, \xk, PTF11rks and SN2012il from 
\cite{11xk}.  They appear to be around $50\%$ lower than \ap,  
and have a similar evolutionary trend. We will return to the discussion and 
model fits in Section\,\ref{disc}.

\begin{figure*}$
\begin{array}{cc}
\includegraphics[angle=270, scale=0.325]{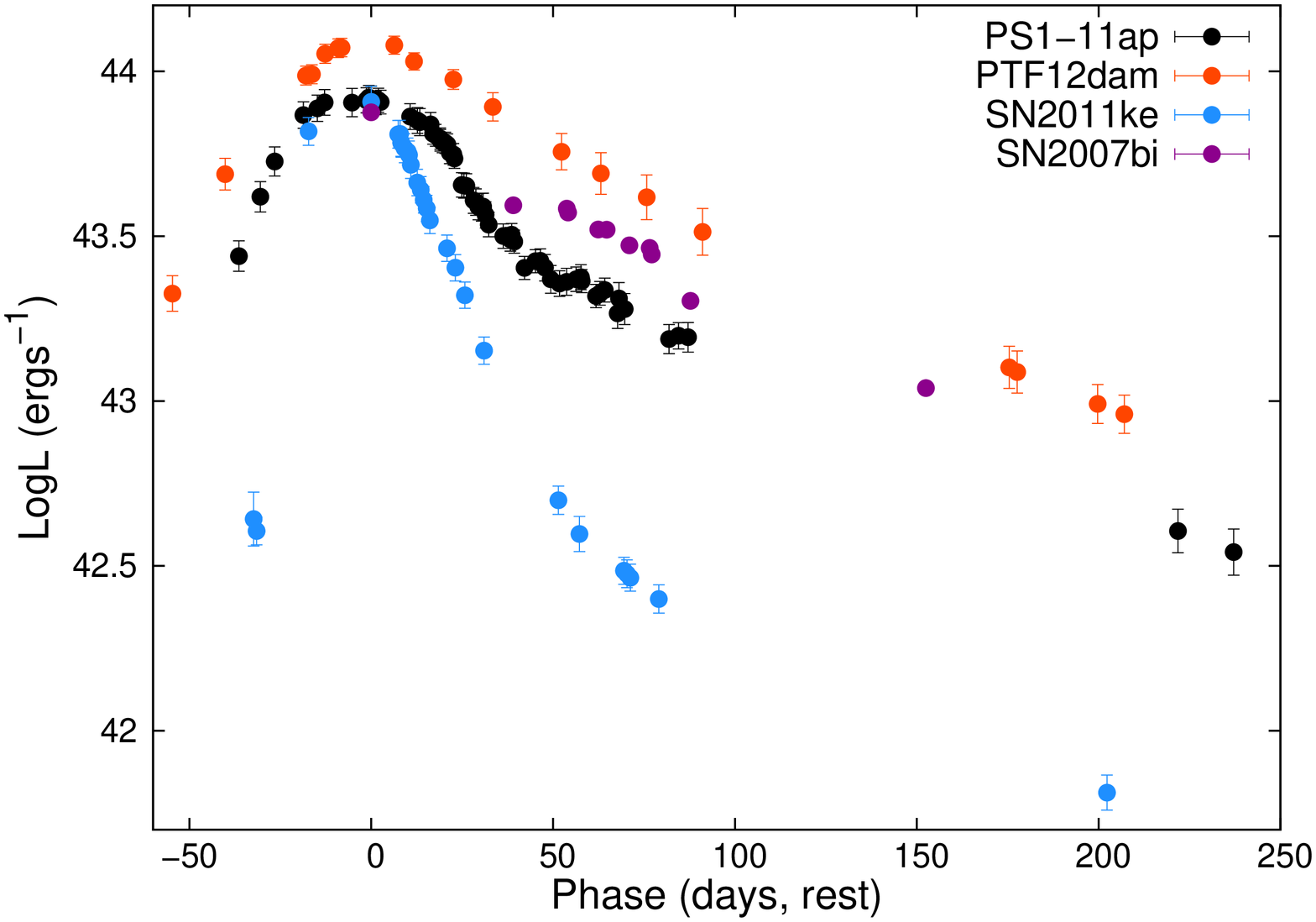} &
\includegraphics[angle=270,scale=0.325]{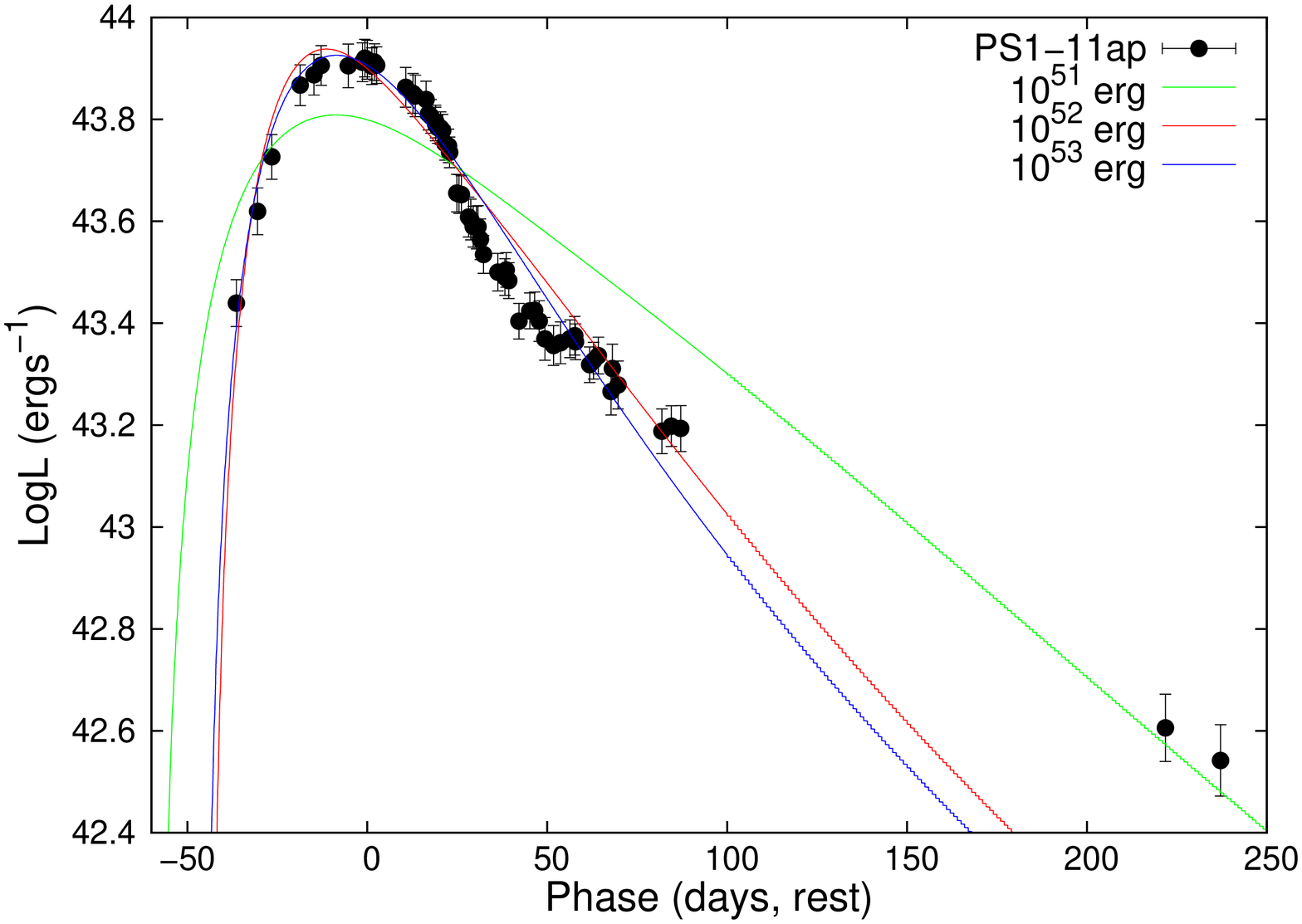} \\
\includegraphics[angle=270,scale=0.325]{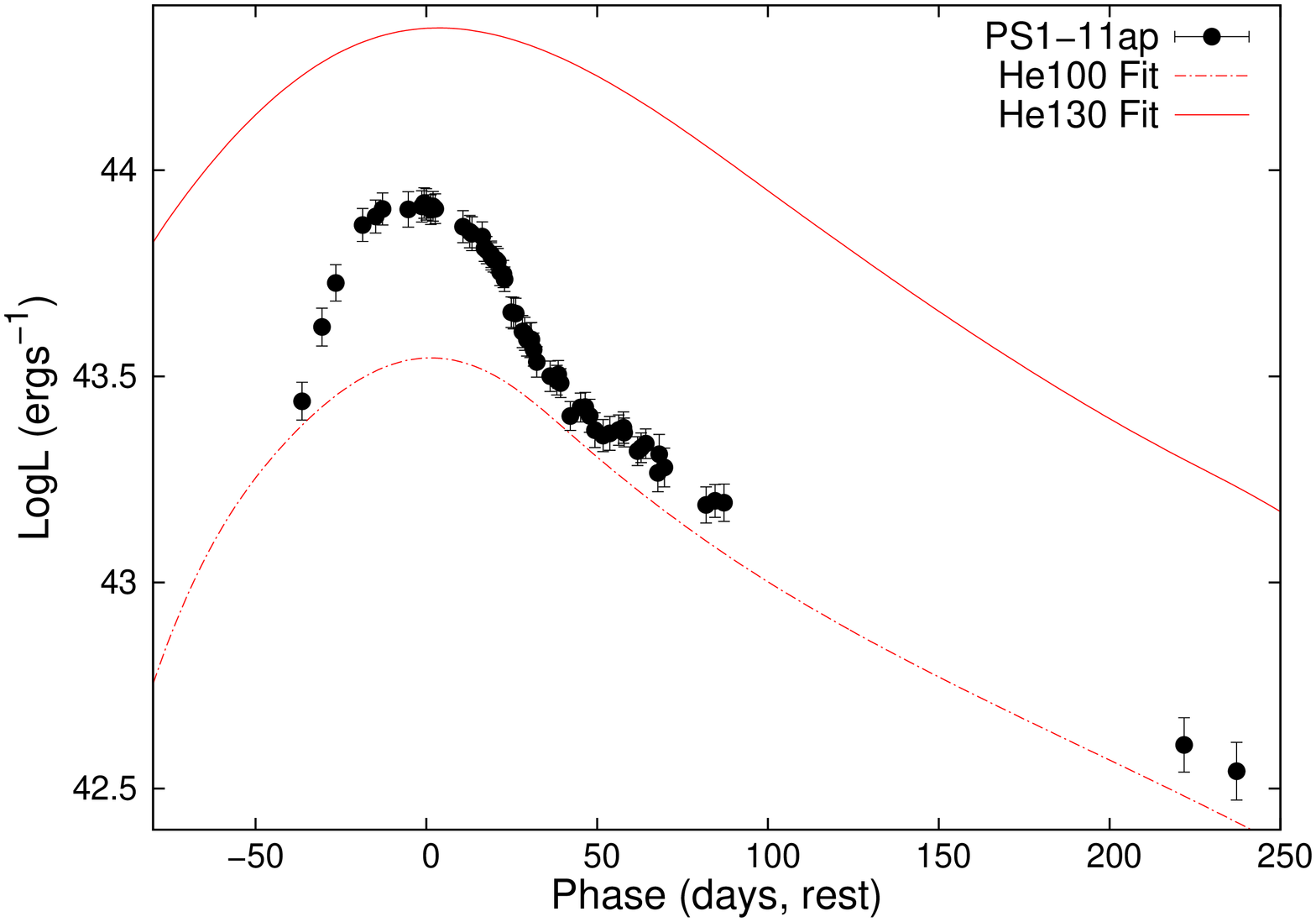} &
\includegraphics[angle=270,scale=0.325]{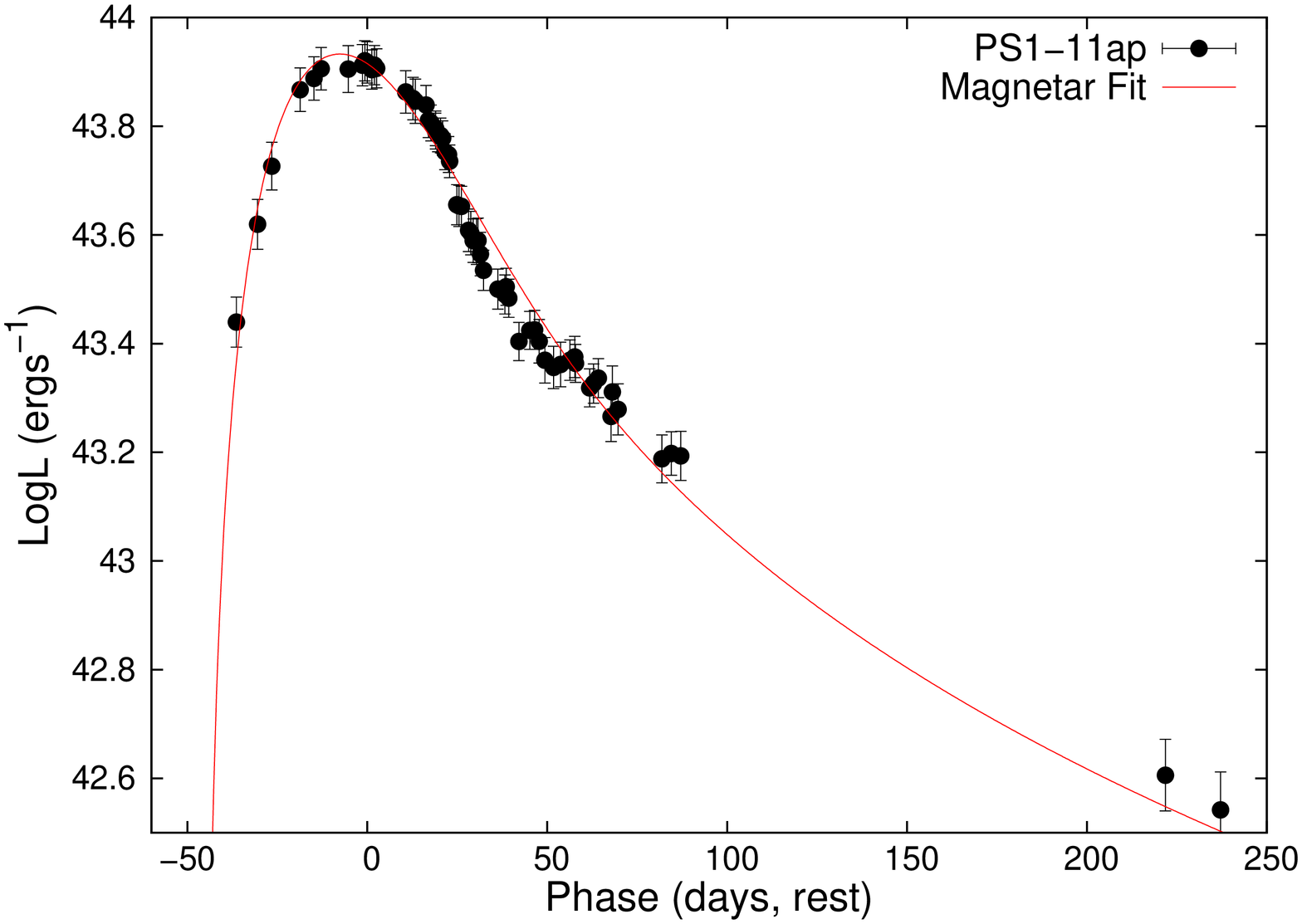} \\
\end{array}$
\caption{A comparison of the \ap\ bolometric light curve with a selection of other SLSNe and with various models.  \emph{({\bf a}) TOP LEFT PANEL}: The bolometric comparison with \dam,  \xk\ and \bi\ (see references in text).  \emph{({\bf b}) TOP RIGHT PANEL}: A comparison with models based on varying amounts of Ni$^{56}$ being synthesised.  \emph{({\bf c}) BOTTOM LEFT PANEL}: How the data compares with two PISNe models with a range of He core masses that should include \ap.  \emph{({\bf d}) BOTTOM RIGHT PANEL}: The best fitting magnetar model for the \ap\ bolometric light curve.}
 \label{fig:11apfits}
\end{figure*}

\section{Discussion}
\label{disc}

\begin{table*}
 \caption{Parameters for the $^{56}$Ni model fits and the magnetar fit shown in Fig.\,\ref{fig:11apfits}.}
 \label{tab:11ap_nimods}
 \begin{tabular}{@{}lccccc}
  \hline
  \hline
Energy (erg) & Ejecta Mass (\msun) & Mass of $^{56}$Ni (\msun) & Rise time (days) & $\chi^2$ \\
 \hline
$10^{51}$ & 7.0 & 6.9 & 58.60 & 145.0 \\
$10^{52}$ & 9.5 & 7.3 & 43.96 & 46.0 \\
$10^{53}$ & 25.0 & 8.2 & 45.69 & 42.0 \\
  \hline
  \hline
Energy (erg) & Ejecta Mass (\msun) & Period (ms) & Rise time (days) & $\chi^2$ \\
&&Magnetic Field (G) &&\\
 \hline
$10^{51}$ & 7.2 & 3.9 & 45.9 & 13.5 \\
&&$2\times10^{14}$&&\\
\hline
 \end{tabular}
 \medskip
\end{table*}

\subsection{Bolometric light curve}

As only \grizy\ photometry was available for \ap, a true bolometric light curve could not be constructed from direct observations alone.  In the galaxy rest frame of \ap\ the rest wavelengths covered by the five \PS\ filters ranges from $2620\AA-6820\AA$.  However photometry in the far UV and the NIR for \dam\ \citep{12dam} are useful supplementary data.  If we assume that \ap\ and \dam\ are powered by the same physical process, which is justified given the similarities in their spectral behaviour and photometric evolution, then we can use the \dam\ SED to complete the \ap\ SED at each of our bolometric epochs.  A full bolometric light curve, which mirrors the epochs of the \ap\ \ips-band data, could then be constructed by integrating the fluxes of the composite \ap\ and \dam\ photometry after the observed magnitudes of each were given a full K-correction.  This data can be found in Table\,\ref{tab:11ap_bollc} in Appendix\,\ref{app:A}.  As seen in Fig.\,\ref{fig:11ap_lc} the \grizy\ coverage for \ap\ is extensive across all epochs but if a filter was not observed for a given night an approximate \gps, \rps, \ips, \zps\ or \yps\ magnitude could be obtained by interpolating the light curves using colour constants from the closest available epochs.

Bolometric data for \bi\ was taken from \cite{07bib} and also amalgamated with \dam\ data to produce a full, composite bolometric light curve which better serves our intentions of comparison than the pseudo-bolometric data presented in the original paper.  A similar method was used in \cite{11xk} to produce the \xk\ bolometric data also used here.  The final \ap\ bolometric light curve, along with a complete bolometric light curve for \dam\ and full, composite bolometric light curves for \xk\ and \bi\ \citep{12dam,11xk,07bib}, is shown in plot (a) of Fig.\,\ref{fig:11apfits}.  As in the case of the absolute magnitude comparison \ap, \dam\ and \bi\ stand apart from the SLSN-Ic \xk.

\subsection{SLSNe models}

The majority of recent publications on SLSNe have focused on three
physical channels to produce the luminosity (see \cite{chom} for a
detailed example). In his recent review of luminous supernovae,
\cite{gal-yam} attempted to collate all the published events into
three classes; SLSNe-II, SLSNe-I and SLSNe-R, and then to associate
each with particular types of physical model. This brief discussion
will be structured into seeing which of these classes, if any, best
fits the observed data for \ap\ and by implication, other objects like
it.

The defining observational feature of the SLSN-II class is the
presence of strong hydrogen emission lines in the spectra often with
multiple velocity components as seen in SN2006gy \citep{06gya,
  06gyb}. This is fairly convincing evidence of interaction between
ejecta from a SN explosion (core-collapse or possibly pulsational
pair-instability) and pre-existing circumstellar material. There is no
clear signature of interaction in any of the SLSN-I or SLSN-Ic events
\citep{10gx,quimb,chom,gal-yam,11xk}. However
\cite{2011ApJ...729L...6C} and \cite{2012ApJ...757..178G} argue that
interaction with a very dense CSM shell of material would not produce
such classic signatures and that this is still a viable model for some
SLSNe. We certainly see no features indicating that \ap\ is
interacting but we will return to this point later. 

The comparative plot in Fig.\,\ref{fig:11apspeccomp2} shows that there
are spectral similarities between objects of the SLSN-Ic class and \ap,
for example the early time blue, almost featureless spectra as seen in \gx\
\citep{10gx,quimb}. 
Where \ap\ does not fit well into the SLSN-Ic class
is in its photometric evolution. This is apparent in both the absolute
magnitude comparison shown in Fig.\,\ref{fig:11apab} and in the
bolometric light curve comparison shown in Fig.\,\ref{fig:11apfits}.
These figures also highlight the similarities between the \ap\ light
curves and the light curves for the SLSNe \bi\ and \dam. The
spectroscopic similarities between these three objects are presented in
Fig.\,\ref{fig:11apspeccomp} suggesting that \ap\ is most similar to
\bi\ and \dam. These three low to moderate redshift SNe stand out as
being quite different in their lightcurve evolution compared to all
other SLSN-Ic presented in the literature so far in that
they fade much more slowly from peak throughout the 200-500 days rest
frame period.

This slow decline in the luminosity of \bi\ has previously thought to
have been driven by the radioactive decay of $3-7$ \msun\ of
$^{56}$Ni \citep{07bia,07bib}, which is converted to $^{56}$Fe via the decay of $^{56}$Co.
This large, but not unphysical \citep{massfea},
amount of $^{56}$Ni has been proposed to have been produced as a
result of a PISN \citep{07bia} or from an upscaled version of the iron
core collapse model used to explain `standard' core-collapse SNe
\citep{07bib,massfeb}.  The latter use the argument that the host
environment of \bi\ is unsuitable for the genesis of stars massive
enough for the pair production mechanism to occur
although \cite{yusof} can now produce model PISN progenitors from moderate metallicity ($\sim0-0.3$Z$_{\odot}$), very massive stars ($100-290$\msun).
An alternative energy source for the SN, in the form of rotational energy being
released from the spin down of a rotating neutron star, has also been
proposed for \bi\ by \cite{magmod} and more recently by \cite{des}.
\cite{magmod} note that their magnetar models may have trouble reproducing
the iron emission lines seen in the nebular phase spectrum of \bi,
which would be expected in the later epochs of the $^{56}$Ni models,
whereas \cite{des} state that the observed post-peak \bi\ spectra do
not show the line blanketing from the iron-group elements and overall
red colours as a result of the cool photospheric temperatures expected
in PISNe.
\cite{yusof} claim that the reduction in angular momentum as a result of the mass loss of these very massive stars as they become WR stars negates the possibility of a magnetar being formed due to a lack of rotational velocity.  The physical production of a magnetar is at present an unclear process but their existence in the Milky Way has been confirmed \citep{2013ApJ...775L..34R}.

If \ap\ is of the same physical  origin as \bi\ do we encounter
the same problems when trying to tie down a single progenitor channel?
The observed characteristics of \ap\ have the same tension with PISN model predictions as \bi\ and \dam.
In Sect.\,\ref{host} we estimated a lower limit on the metallicity of the host galaxy to be $\sim0.2$Z$_{\odot}$ which is similar 
to that found for the host of SN2007bi \citep{07bib} and PTF12dam (Chen
et al. in prep.).  Within the error margins this value is low enough that 
the progenitor star of the SN could evolve a carbon-oxygen core of 
$\sim 60 - 130\msun$ for pair 
production to occur \citep{yusof}.
Hence it is theoretically plausible that a PISN progenitor could be produced at the estimated metallicity of Z $\gtrsim0.2$Z$_{\odot}$.
In contrast to this the colour evolution of \ap\ shows a redder trend
than that of \bi\ but still falls short of the $B-R=1.47$ value
expected at $\sim50$ days from PISN model spectra \citep{des}.  The
$\gps-\rps$ value for \ap\ at this epoch is $\sim0.23$.  The
differences in the redshift and the observed filters used for each
object makes a direct comparison difficult. Fig.\,\ref{fig:11apfits}(c) also shows how poorly PISN models match the well sampled \ap\ photometry.

Plots (b) and (d) of Fig.\,\ref{fig:11apfits} show two further types of model fit to the derived bolometric light curve for \ap, which we discuss in the following sections. 

\subsection{Model explosions powered by $^{56}$Ni}
\label{mods}

\cite{arnett} produced model Type I SNe light curves using the
radioactive decay of $^{56}$Ni as a power source in a homologously
expanding ejecta where radiation pressure is dominant and a constant
opacity is assumed.  The models used in Fig.\,\ref{fig:11apfits}(b) are
based on this semi-analytic treatment and give us a photometric
evolution using the decay of $^{56}$Ni to $^{56}$Co and eventually
$^{56}$Fe (half lives of 6.08 days and 77.23 days respectively) to power
the SN.  Further details of this implementation are given in \cite{valenti} and \cite{11xk}. 
The model has four free input parameters; the energy of the
explosion, the total ejecta
mass (M$_{ej}$), the mass of $^{56}$Ni and the explosion date . These parameters determine the
light curve shape including the peak luminosity, rise time and decay time.
\cite{massfea} has discussed physically plausible upper limits
on $^{56}$Ni that can be produced in massive SNe with a limit on the ratio of $^{56}$Ni to M$_{ej}$ of 0.2.

Models that use $^{56}$Ni as a power source for the very large
luminosities of SLSNe can account for the light curve rise time and peak 
reasonably well, but they fail to match the observed decay out to 250
days.  Model values are listed in Table\,\ref{tab:11ap_nimods}.
Note that there is considerable degeneracy in the energy-mass combination for these models, hence the inclusion of fits for three fixed energies as opposed to a single best fit over all four parameters.
The models with 10$^{51}$ and 10$^{52}$ ergs are not physically
realistic as they are almost 99\% and 77\% Ni by mass
fraction.
The observed spectra are also not Fe dominated as one would
expect. \cite{chom} has previously found that the ratios of the
mass of $^{56}$Ni to M$_{ej}$ were not plausible in the case of
PS1-10ky and PS1-10awh and \cite{12dam} found 
similar results for \dam\ as we do here. This is not surprising given the
similarity of the SNe.
The best fit to the \ap\ bolometric light curve (formally the lowest $\chi^{2}$) is given by the
highest energy model ($10^{53}$ ergs).  However the model fits do not
match the late photometric points at 250 days and there is no evidence,
as yet, of strong Fe\,{\sc ii} emission in the pseudo-nebular spectrum
 at ($\sim200$ days) that we would expect with 8\msun\ of $^{56}$Ni \citep{gypisn}.

Fig.\,\ref{fig:11apfits}(c) illustrates two PISN models from
\cite{pisnmod} for different He-core masses of the progenitor star
core.  Although the peak luminosity is not specifically matched for the \ap\ data
by either of the models, a model of around 115\msun\ would likely fit
the peak. But more importantly,  it can clearly be seen that neither model
fits the overall evolution of the data.
\ap\ has a shorter and brighter diffusion phase ($\sim$-50 to 50 days) compared to the models suggesting a lower M$_{ej}$.
\cite{12dam} find a similar problem when comparing the \dam\ light
curve with PISNe models, where the earlier sampled rise time phase
also does not match the predicted gradual increase in brightness.
Line blanketing expected from iron group elements in the UV range of the model PISNe
spectra is also not present in the blue \dam\ spectra.  The higher
redshift of \ap\ allows us to probe further into the restframe UV with
optical spectra and does show broad absorption from Mg\,{\sc ii} but
the overall SED is still too blue for the predicted model spectra.
\cite{des} find a similar inconsistency with the \bi\ spectra.
Hence no single set of the input
parameters for the $^{56}$Ni models, detailed in
Table\,\ref{tab:11ap_nimods}, or the He-core PISN models of
\cite{pisnmod} accurately mimic the very well sampled \ap\ data.

\subsection{Model explosions powered by magnetar spin down}

The final fit presented, Fig.\,\ref{fig:11apfits}(d), shows the
light curve evolution of a supernova arising from the extra input of
energy from a magnetic neutron star which is initially rapidly
rotating. This object, often referred to as a magnetar, spins down 
due to magnetic braking and energises the expanding 
remnant through some coupling mechanism to power a more luminous SN 
\citep{pulsara,pulsarb,magmod}.

We use a paramaterised semi-analytic approach which takes the 
energy from a spinning down magnetar and inputs this into the Arnett 
diffusion solutions \citep{arnett}. This is  similar to
the approach detailed in Section\,\ref{mods} with the $^{56}$Ni power
source replaced with magnetar luminosity.  Full details can be 
found in \cite{11xk} (see the Appendix).
 The  luminosity injected by the magnetar 
depends on the magnetic field strength, $B$, and the initial 
period, $P_i$. The explosion energy, ejecta mass and explosion date are still free
parameters in this approach, hence we have 
 increased  the number of free input parameters to 5, as compared to 4
 in the $^{56}$Ni models.  The range of different light curves that can be
produced with physically realistic values for the M$_{ej}$ is also
much more diverse as they are now uncoupled from the power source (the
magnetar).  In addition to this there is also uncertainty over the
conversion of energy from the X-rays and gamma rays produced by the
magnetar emission into the observed luminosity of the SN \citep{uncert} and full
trapping of these gamma rays must be assumed to successfully reproduce
the massive luminosity observed.  However the large and metal
dominated ejecta supports the trapping assumptions.

\cite{magmod} successfully used magnetar models to fit the observed
\bi\ photometry.  Their best fit model had values for a supernova with
M$_{ej}=20$ \msun\ that formed a magnetar with a magnetic field,
$B=2\times10^{14}$ G, and a period, $P_{i}=2.5$ ms.  The best fit for
\ap\ has the following comparable parameters; M$_{ej}=7.2$ \msun,
$B=2\times10^{14}$ G and $P_{i}=3.9$ ms.  
The minimum period that a rotating neutron star can have is $\sim1$ ms \citep{magmod} and our model fit sits comfortably above this boundary.
The differences in the expected temperature evolution of a $^{56}$Ni
powered SLSN and the actual temperature evolution of \bi\ was a major
argument against the PISN scenario in \cite{des} and more
recently in \cite{12dam}.   The
lower right hand plot in Fig.\,\ref{fig:11apbb} shows how the expected
temperature evolution of our magnetar model matches the temperature
evolution of \ap.  The values deduced for the data seem to
exceed those of the model but, although the difference is around 10\%,
the error bars overlap most of the points and the shape of the
T$_{eff}$ curve produced by the model matches the data reasonably
well. It certainly appears that the temperature evolution is much
better matched with the magnetar model than the PISN model. 

This presents an interesting issue of unifying the whole class of SLSNe.  \cite{11xk} use magnetar models
to explain the SLSNe-Ic in their dataset (which includes \xk).  The
differences in the photometric evolution of objects in the SLSN-Ic
class and objects like \bi\ or \ap\ has previously suggested a
completely separate physical mechanism powering the SNe
\citep{gal-yam,quimb}.  Both the absolute magnitude comparison and the
bolometric light curve comparison presented in this paper show
distinctly different trends for the two supposed classes.  However the
ability to fit the same physical model to the bolometric light curves
of both classes and the spectral comparison presented here in
Fig.\,\ref{fig:11apspeccomp2} offers evidence towards one analogous
physical mechanism. We suggest that all of these are more sensibly
labelled as SLSN-Ic, as they are clearly ``super''-luminous and are 
type Ic in the standard SN classification schemes. With the added
dense spectral coverage from \cite{12dam} and now with \ap, 
the spectral 
evolution is more homogeneous than previously thought. 
More importantly it
appears that one single physical mechanism could plausibly power the 
luminosity of all of these SLSNe-Ic.  The differences in the 
light curve shapes may arise mostly from differences in ejecta mass, initial magnetar spin rate and magnetic field
or from the fact that this large number of free parameters produces an unrealistic range of possible model outputs.

A possibility that should be mentioned for completeness is that the
observed photometric properties of \ap\ could be reproduced by having
both a magnetar and some radioactive decay as power sources.  This is
explored in greater detail by \cite{11xk}.

\subsection{Model explosions powered by interaction} 

Another possible physical mechanism for producing light curves like
those of SLSNe is the shock breakout and interaction model described by
\cite{2011ApJ...729L...6C}.  In this model kinetic energy from the SN
is converted into radiation through interaction with an optically
thick circumstellar medium (CSM).  The predicted photometric and
spectroscopic properties then depend on the relationship of the cutoff
distance of the CSM ($r_\omega$) and the diffusion radius of the SN
($r_d$).  Recently, \cite{2012ApJ...757..178G} used models based on
this idea to reproduce the light curves of the SLSN-Ic \gx\ and the
SLSN-II SN2006gy.  By setting $r_\omega>r_d$ not only can the light
curves of SLSNe-II be reproduced but the broad \ha\ feature as well.
Setting $r_\omega \lesssim r_d$ the light curve shape and distinctive
broad `W'-shaped feature from O\,{\sc ii} in the spectra of SLSNe-I
are predicted \citep{2011ApJ...729L...6C}.

\cite{chatz} also use CSM models to fit SLSNe data.  They produce fits to the SLSN-I, SLSN-II classes and to \bi.  Model fits using radiative decay and magnetars as energy sources are investigated as well.  Regarding \bi, \cite{chatz} reach the conclusion that a PISN is not a satisfactory progenitor channel for the SN.  The paper focuses on a CSM explanation however their magnetar model is clearly the best fit to the observed data, especially when the late-time points (at $\sim300$ days) are taken into account.  It is noted that the CSM models used have the largest number of free parameters and that external factors such as the ejecta geometry and clumping of the CSM can also have a large effect on the observed properties of any observed SN.

\section{Conclusions}

A single explosion with a radiated energy of $10^{51}$ ergs and an
overall light curve evolution that lasted $>200$ days was observed for
\ap.  After comparing the obtained \grizy\ light curves and multiple
spectra with literature data from various SLSNe there can be little
doubt that \ap\ is yet another member of this emerging class of SN.
The derived redshift of $z=0.524$ also makes this SLSN fit nicely into
the intermediate redshift region between the reasonably high number of objects of
this nature found at $z\simeq0.1$ and $z\simeq2$.  In particular,
similarities with \bi\ and \dam\ suggest that \ap\ has the same
physical origins as these rare, slowly declining transients.  The data gathered
here do not shed any light on the inconsistencies
associated with the PISN explanation for \bi\ and in fact our analysis points further away from this explanation.  The
lack of strong nebular emission lines of iron in the late time spectra also seems to
negate the possible massive core collapse scenario.  The most
consistent conclusion is that \ap\ is powered by the spin down of a
magnetar.  The magnetar driven model provides a good  fit to the
extensive photometric  data and is consistent with the colour evolution of our spectral
series.

Of interest is the spectral similarities between \ap\ and the SLSN-Ic
class when taken in the context of recent work that successfully fits
magnetar models to light curves of various objects that fall into this
category.  Whether there is a physical link between these apparently
separate classes or whether this is simply a shortcoming of the overly
flexible magnetar models remains to be seen until more objects with
similar properties to \ap\ are discovered and more in-depth magnetar
models have been produced.

{\it Facilities:} PS1 (GPC1)

{\small{ \textit{Acknowledgements}. 

    The Pan-STARRS1 Surveys (PS1) have been made possible through
    contributions of the Institute for Astronomy, the University of
    Hawaii, the Pan- STARRS Project Office, the Max-Planck Society and
    its participating institutes, the Max Planck Institute for
    Astronomy, Heidelberg and the Max Planck Institute for
    Extraterrestrial Physics, Garching, The Johns Hopkins University,
    Durham University, the University of Edinburgh, Queen's University
    Belfast, the Harvard- Smithsonian Center for Astrophysics, the Las
    Cumbres Observatory Global Telescope Network Incorporated, the
    National Central University of Taiwan, the Space Telescope Science
    Institute, the National Aeronautics and Space Administration under
    Grant No. NNX08AR22G issued through the Planetary Science Division
    of the NASA Science Mission Directorate, the National Science
    Foundation under Grant No. AST-1238877, and the University of
    Maryland. S.J.S. acknowledges funding from the European Research
    Council under the European Union's Seventh Framework Programme
    (FP7/2007-2013)/ERC Grant agreement no [291222] (PI: S.J. Smartt).
    J.T. acknowledges support for this work provided by National
    Science Foundation grant AST-1009749. R.P.Kir. thanks the National
    Science Foundation for AST-1211196. This work is based on
    observations made with the following telescopes: William Herschel
    Telescope (operated by the Isaac Netwon Group), Nordic Optical
    Telescope (operated by the Nordic Optical Telescope Scientific
    Association)) and Liverpool Telescope (operated by Liverpool John
    Moores University with financial support from the UK Science and
    Technology Facilities Council), all in the Spanish Observatorio
    del Roque de los Muchachos of the Instituto de Astrofísica de
    Canarias, in the island of La Palma; the Gemini Observatory, which is operated by the 
    Association of Universities for Research in Astronomy, Inc., under a cooperative agreement 
    with the NSF on behalf of the Gemini partnership: the National Science Foundation 
    (United States), the National Research Council (Canada), CONICYT (Chile), the Australian 
    Research Council (Australia), Minist\'{e}rio da Ci\^{e}ncia, Tecnologia e Inova\c{c}\~{a}o 
    (Brazil) and Ministerio de Ciencia, Tecnolog\'{i}a e Innovaci\'{o}n Productiva (Argentina).  Some observations reported here were obtained at the MMT Observatory, a joint facility of the Smithsonian Institution and the University of Arizona.  Support for SR was provided by NASA through Hubble Fellowship grant \#HST-HF-51312.01 awarded by the Space Telescope Science Institute, which is operated by the Association of Universities for Research in Astronomy, Inc., for NASA, under contract NAS 5-26555.
. We wish to thank Dan Kasen and
    Luc Dessart for sending us their model data, and Roger Chevalier for discussion.}

\appendix

\section{\ap\ photometry}
\label{app:A}

Table\,\ref{tab:11ap_phot} shows the non K-corrected \grizy\ photometry for \ap\ and Table\,\ref{tab:11ap_kc} the K-correction values for the 4 different epochs used.
The derived bolometric data for \ap\ is shown in Table\,\ref{tab:11ap_bollc}.

\begin{landscape}

\begin{table}
\tiny
 \caption{Observed photometry of \ap, without K-corrections.  The LT magnitudes have been converted to the \PS\ system as discussed in Section\,2.2.  The MJD and phase of any observations from December 2011 onwards represent the midpoints of the co-added frames, hence the uncertainties given.}
 \label{tab:11ap_phot}
 \begin{tabular}{cccccc | cccccc | cccccc}
  \hline
  \hline
Date & MJD & Phase (days, rest) & M (dm) & Filter & Telescope & Date & MJD & Phase (days, rest) & M (dm) & Filter & Telescope & Date & MJD & Phase (days, rest) & M (dm) & Filter & Telescope \\
    \hline
31/12/2010 & 55561.60 & -35.04 & 21.50 (0.04) & \ips & PS1 & 22/03/2011 & 55642.96 & 18.35 & 20.33 (0.02) & \emph{r} & LT &	23/04/2011& 55674.33 & 38.93 & 20.97 (0.03) & \rps & PS1 \\
02/01/2011 & 55563.61 & -33.72 & 21.64 (0.10) & \gps & PS1 & 22/03/2011 & 55642.97 & 18.35 & 20.36 (0.03) & \emph{i} & LT &	23/04/2011& 55674.94 & 39.33 & 21.60 (0.07) & \emph{g} & LT \\
02/01/2011 & 55563.63 & -33.71 & 21.31 (0.07) & \rps & PS1 & 24/03/2011 & 55644.01 & 19.03 & 20.82 (0.04) & \emph{g} & LT &	 24/04/2011& 55675.33 & 39.58 & 20.79 (0.02) & \ips & PS1 \\
09/01/2011 & 55570.55 & -29.16 & 21.05 (0.05) & \ips & PS1 & 24/03/2011 & 55644.01 & 19.04 & 20.35 (0.03) & \emph{r} & LT &	24/04/2011& 55675.89 & 39.95 & 21.67 (0.05) & \emph{g} & LT \\
11/01/2011 & 55572.58 & -27.83 & 21.11 (0.07) & \gps & PS1 & 24/03/2011 & 55644.03 & 19.05 & 20.38 (0.04) & \emph{i} & LT &	24/04/2011& 55675.89 & 39.96 & 20.87 (0.04) & \emph{r} & LT \\
11/01/2011 & 55572.60 & -27.82 & 20.93 (0.05) & \rps & PS1 & 24/03/2011 & 55644.03 & 19.05 & 20.38 (0.07) & \emph{z} & LT &	24/04/2011& 55675.90 & 39.96 & 20.67 (0.04) & \emph{i} & LT \\
15/01/2011 & 55576.62 & -25.18 & 20.75 (0.04) & \ips & PS1 & 25/03/2011 & 55645.45 & 19.98 & 20.37 (0.02) & \ips & PS1 &24/04/2011& 55675.90 & 39.96 & 20.61 (0.09) & \emph{z} & LT \\
21/01/2011 & 55582.44 & -21.36 & 20.76 (0.16) & \yps & PS1 & 25/03/2011 & 55645.98 & 20.33 & 20.88 (0.04) & \emph{g} & LT &25/04/2011& 55676.34 & 40.25 & 20.81 (0.06) & \zps & PS1 \\
22/01/2011 & 55583.52 & -20.66 & 20.78 (0.04) & \zps & PS1 &  25/03/2011 & 55645.98 & 20.33 & 20.32 (0.03) & \emph{r} & LT &25/04/2011& 55676.96 & 40.66 & 21.74 (0.06) & \emph{g} & LT \\
23/01/2011 & 55584.56 & -19.97 & 20.65 (0.06) & \gps & PS1 & 25/03/2011 & 55645.99 & 20.34 & 20.43 (0.04) & \emph{i} & LT &	25/04/2011& 55676.97 & 40.66 & 20.97 (0.04) & \emph{r} & LT \\
23/01/2011 & 55584.57 & -19.96 & 20.51 (0.04) & \rps & PS1 & 26/03/2011 & 55646.00 & 20.34 & 20.39 (0.10) & \emph{z} & LT &	25/04/2011& 55676.97 & 40.66 & 20.61 (0.04) & \emph{i} & LT \\
25/01/2011 & 55586.61 & -18.63 & 20.70 (0.03) & \zps & PS1 & 26/03/2011 & 55646.43 & 20.63 & 20.52 (0.03) & \zps & PS1 &26/04/2011& 55677.38 & 40.93 & 21.84 (0.04) & \gps & PS1 \\
26/01/2011 & 55587.63 & -17.96 & 20.58 (0.03) & \gps & PS1 & 26/03/2011 & 55646.94 & 20.96 & 20.91 (0.03) & \emph{g} & LT &26/04/2011& 55677.39 & 40.94 & 21.07 (0.03) & \rps & PS1 \\
26/01/2011& 55587.64 & -17.95 & 20.40 (0.02) & \rps & PS1 &	 26/03/2011 & 55646.95 & 20.96 & 20.39 (0.02) & \emph{r} & LT &	29/04/2011& 55680.26 & 42.82 & 22.06 (0.07) & \gps & PS1 \\
27/01/2011& 55588.64 & -17.30 & 20.46 (0.02) & \ips & PS1 &	 26/03/2011 & 55646.95 & 20.97 & 20.39 (0.03) & \emph{i} & LT &	29/04/2011& 55680.27 & 42.83 & 21.11 (0.04) & \rps & PS1 \\
28/01/2011& 55589.56 & -16.69 & 20.63 (0.03) & \zps & PS1 & 26/03/2011 & 55646.96 & 20.97 & 20.41 (0.08) & \emph{z} & LT &	30/04/2011& 55681.27 & 43.48 & 20.81 (0.03) & \ips & PS1 \\
29/01/2011& 55590.57 & -16.03 & 20.55 (0.03) & \gps & PS1 & 28/03/2011 & 55648.09 & 21.72 & 20.89 (0.03) & \emph{g} & LT &	 01/05/2011 & 55682.25 & 44.13 & 20.82 (0.04) & \zps & PS1 \\
29/01/2011& 55590.58 & -16.02 & 20.45 (0.03) & \rps & PS1 &	 28/03/2011 & 55648.10 & 21.72 & 20.41 (0.03) & \emph{r} & LT & 04/05/2011 & 55685.89 & 46.52 & 21.96 (0.05) & \emph{g} & LT \\
31/01/2011& 55592.58 & -14.71 & 20.58 (0.03) & \zps & PS1 & 28/03/2011 & 55648.11 & 21.72 & 20.42 (0.04) & \emph{i} & LT & 04/05/2011 & 55685.90 & 46.52 & 21.09 (0.05) & \emph{r} & LT \\
01/02/2011& 55593.61 & -14.04 & 20.55 (0.03) & \gps & PS1 & 28/03/2011 & 55648.11 & 21.73 & 20.42 (0.08) & \emph{z} & LT & 04/05/2011 & 55685.90 & 46.52 & 20.70 (0.05) & \emph{i} & LT \\
01/02/2011& 55593.62 & -14.03 & 20.36 (0.02) & \rps & PS1 &	 28/03/2011 & 55648.98 & 22.30 & 20.94 (0.03) & \emph{g} & LT &	 05/05/2011 & 55686.92 & 47.19 & 21.96 (0.06) & \emph{g} & LT \\
02/02/2011& 55594.45 & -13.48 & 20.38 (0.02) & \ips & PS1 &	 28/03/2011 & 55648.98 & 22.30 & 20.37 (0.03) & \emph{r} & LT & 05/05/2011 & 55686.93 & 47.20 & 21.06 (0.06) & \emph{r} & LT \\
03/02/2011& 55595.53 & -12.78 & 20.54 (0.02) & \zps & PS1 & 28/03/2011 & 55648.99 & 22.30 & 20.44 (0.04) & \emph{i} & LT &	 06/05/2011 & 55687.89 & 47.83 & 21.92 (0.06) & \emph{g} & LT \\
04/02/2011& 55596.51 & -12.13 & 20.48 (0.03) & \gps & PS1 & 28/03/2011 & 55648.99 & 22.31 & 20.38 (0.09) & \emph{z} & LT &	 06/05/2011 & 55687.90 & 47.83 & 21.01 (0.05) & \emph{r} & LT \\
04/02/2011& 55596.53 & -12.12 & 20.26 (0.02) & \rps & PS1 &	 29/03/2011 & 55649.50 & 22.64 & 20.55 (0.06) & \zps & PS1 & 06/05/2011 & 55687.90 & 47.84 & 20.81 (0.07) & \emph{i} & LT \\
05/02/2011& 55597.51 & -11.48 & 20.35 (0.02) & \ips & PS1 &	 29/03/2011 & 55649.98 & 22.95 & 21.01 (0.04) & \emph{g} & LT &	 08/05/2011 & 55689.89 & 49.14 & 22.07 (0.12) & \emph{g} & LT \\
10/02/2011& 55602.43 & -8.25 & 20.37 (0.04) & \gps & PS1 &	 29/03/2011 & 55649.99 & 22.96 & 20.49 (0.04) & \emph{r} & LT &	 08/05/2011 & 55689.90 & 49.15 & 21.07 (0.07) & \emph{r} & LT \\
10/02/2011& 55602.44 & -8.24 & 20.28 (0.03) & \rps & PS1 & 29/03/2011 & 55649.99 & 22.96 & 20.38 (0.04) & \emph{i} & LT &	 08/05/2011 & 55689.91 & 49.15 & 20.77 (0.07) & \emph{i} & LT \\
16/02/2011& 55608.43 & -4.31 & 20.41 (0.10) & \yps & PS1 &	 30/03/2011 & 55650.00 & 22.96 & 20.33 (0.08) & \emph{z} & LT &	 09/05/2011 & 55690.97 & 49.85 & 22.03 (0.16) & \emph{g} & LT \\
17/02/2011& 55609.02 & -3.92 & 20.54 (0.06) & \emph{g} & LT & 30/03/2011& 55650.48 & 23.28 & 20.99 (0.04) & \gps & PS1 & 09/05/2011 & 55690.98 & 49.85 & 21.23 (0.08) & \emph{r} & LT \\
17/02/2011& 55609.03 & -3.92 & 20.21 (0.05) & \emph{r} & LT & 30/03/2011& 55650.50 & 23.29 & 20.49 (0.03) & \rps & PS1 & 11/05/2011 & 55692.28 & 50.71 & 22.23 (0.13) & \gps & PS1 \\
17/02/2011& 55609.04 & -3.91 & 20.25 (0.03) & \emph{i} & LT & 30/03/2011& 55650.99 & 23.61 & 21.02 (0.05) & \emph{g} & LT & 11/05/2011 & 55692.29 & 50.72 & 21.27 (0.05) & \rps & PS1 \\
18/02/2011& 55610.49 & -2.96 & 20.46 (0.06) & \yps & PS1 & 30/03/2011& 55650.99 & 23.62 & 20.53 (0.04) & \emph{r} & LT & 11/05/2011 & 55692.93 & 51.14 & 21.24 (0.13) & \emph{r} & LT \\
19/02/2011& 55611.03 & -2.61 & 20.46 (0.09) & \emph{g} & LT & 31/03/2011& 55651.47 & 23.93 & 20.43 (0.02) & \ips & PS1 & 11/05/2011 & 55692.94 & 51.14 & 20.75 (0.09) & \emph{i} & LT \\
19/02/2011& 55611.04 & -2.60 & 20.13 (0.05) & \emph{r} & LT & 31/03/2011& 55651.94 & 24.24 & 21.02 (0.03) & \emph{g} & LT & 13/05/2011 & 55694.28 & 52.02 & 20.90 (0.07) & \zps & PS1 \\
19/02/2011& 55611.05 & -2.59 & 20.34 (0.08) & \emph{z} & LT & 31/03/2011& 55651.95 & 24.24 & 20.53 (0.04) & \emph{r} & LT & 14/05/2011 & 55695.94 & 53.11 & 20.85 (0.09) & \emph{i} & LT \\
21/02/2011& 55613.08 & -1.26 & 20.43 (0.09) & \emph{g} & LT & 31/03/2011& 55651.95 & 24.25 & 20.52 (0.05) & \emph{i} & LT & 16/05/2011 & 55697.30 & 54.00 & 20.78 (0.08) & \yps & PS1 \\
21/02/2011& 55613.09 & -1.25 & 20.11 (0.05) & \emph{r} & LT & 02/04/2011& 55653.48 & 25.25 & 21.07 (0.04) & \gps & PS1 & 17/05/2011 & 55698.90 & 55.06 & 20.80 (0.13) & \emph{i} & LT \\
21/02/2011& 55613.44 & -1.02 & 20.38 (0.02) & \zps & PS1 & 02/04/2011& 55653.49 & 25.26 & 20.52 (0.02) & \rps & PS1 & 21/05/2011 & 55702.89 & 57.67 & 22.30 (0.08) & \emph{g} & LT \\
23/02/2011& 55615.02 & 0.02 & 20.49 (0.04) & \emph{g} & LT & 03/04/2011& 55654.94 & 26.21 & 21.11 (0.04) & \emph{g} & LT & 21/05/2011 & 55702.90 & 57.68 & 21.32 (0.06) & \emph{r} & LT \\
23/02/2011& 55615.03 & 0.02 & 20.20 (0.03) & \emph{r} & LT & 03/04/2011& 55654.95 & 26.21 & 20.53 (0.03) & \emph{r} & LT &	 21/05/2011 & 55702.91 & 57.68 & 20.80 (0.05) & \emph{i} & LT \\
23/02/2011& 55615.05 & 0.03 & 20.46 (0.09) & \emph{z} & LT & 03/04/2011& 55654.95 & 26.21 & 20.58 (0.05) & \emph{i} & LT &	 22/05/2011 & 55703.28 & 57.93 & 20.83 (0.05) & \zps & PS1 \\
23/02/2011& 55615.50 & 0.33 & 20.20 (0.03) & \ips & PS1 & 04/04/2011& 55655.94 & 26.86 & 21.18 (0.05) & \emph{g} & LT & 23/05/2011 & 55704.97 & 59.04 & 22.17 (0.06) & \emph{g} & LT \\
24/02/2011& 55616.03 & 0.68 & 20.47 (0.02) & \emph{g} & LT & 04/04/2011& 55655.94 & 26.87 & 20.48 (0.04) & \emph{r} & LT & 23/05/2011 & 55704.98 & 59.04 & 21.37 (0.06) & \emph{r} & LT \\
24/02/2011& 55616.04 & 0.68 & 20.15 (0.02) & \emph{r} & LT & 04/04/2011& 55655.95 & 26.87 & 20.53 (0.06) & \emph{i} & LT &	 23/05/2011 & 55704.99 & 59.05 & 20.84 (0.11) & \emph{i} & LT \\
24/02/2011& 55616.04 & 0.68 & 20.27 (0.02) & \emph{i} & LT & 05/04/2011& 55656.96 & 27.53 & 21.14 (0.05) & \emph{g} & LT &	 24/05/2011 & 55705.26 & 59.23 & 20.93 (0.03) & \ips & PS1 \\
24/02/2011& 55616.05 & 0.69 & 20.24 (0.06) & \emph{z} & LT & 05/04/2011& 55656.96 & 27.54 & 20.57 (0.04) & \emph{r} & LT &	 25/05/2011 & 55706.26 & 59.88 & 20.99 (0.06) & \zps & PS1 \\
25/02/2011& 55617.06 & 1.35 & 20.50 (0.04) & \emph{g} & LT & 05/04/2011& 55656.97 & 27.54 & 20.49 (0.05) & \emph{i} & LT &	 29/05/2011 & 55710.29 & 62.52 & 22.37 (0.06) & \gps & PS1 \\
25/02/2011& 55617.07 & 1.36 & 20.18 (0.02) & \emph{r} & LT & 05/04/2011& 55656.97 & 27.54 & 20.41 (0.08) & \emph{z} & LT &	 29/05/2011 & 55710.30 & 62.53 & 21.38 (0.03) & \rps & PS1 \\
25/02/2011& 55617.08 & 1.36 & 20.18 (0.02) & \emph{i} & LT & 06/04/2011& 55657.94 & 28.18 & 21.30 (0.06) & \emph{g} & LT &	 30/05/2011 & 55711.29 & 63.18 & 21.04 (0.05) & \ips & PS1 \\
25/02/2011& 55617.09 & 1.37 & 20.22 (0.07) & \emph{z} & LT & 06/04/2011& 55657.95 & 28.18 & 20.63 (0.04) & \emph{r} & LT &	 31/05/2011 & 55712.26 & 63.82 & 21.02 (0.03) & \zps & PS1 \\
27/02/2011& 55619.02 & 2.64 & 20.53 (0.02) & \emph{g} & LT & 08/04/2011& 55659.94 & 29.49 & 21.29 (0.05) & \emph{g} & LT & 31/05/2011 & 55712.89 & 64.23 & 22.31 (0.07) & \emph{g} & LT \\
27/02/2011& 55619.03 & 2.64 & 20.16 (0.02) & \emph{r} & LT & 08/04/2011& 55659.95 & 29.49 & 20.70 (0.04) & \emph{r} & LT &	 31/05/2011 & 55712.91 & 64.24 & 21.45 (0.05) & \emph{r} & LT \\
27/02/2011& 55619.04 & 2.65 & 20.22 (0.03) & \emph{i} & LT & 08/04/2011& 55659.95 & 29.50 & 20.52 (0.07) & \emph{i} & LT &	 31/05/2011 & 55712.91 & 64.25 & 20.98 (0.06) & \emph{i} & LT \\
27/02/2011& 55619.05 & 2.66 & 20.52 (0.10) & \emph{z} & LT & 09/04/2011& 55660.94 & 30.14 & 21.41 (0.07) & \emph{g} & LT & 02/06/2011 & 55714.95 & 65.58 & 22.33 (0.07) & \emph{g} & LT \\
28/02/2011& 55620.04 & 3.31 & 20.50 (0.02) & \emph{g} & LT & 09/04/2011& 55660.94 & 30.15 & 20.62 (0.04) & \emph{r} & LT & 02/06/2011 & 55714.96 & 65.59 & 21.36 (0.05) & \emph{r} & LT \\
28/02/2011& 55620.06 & 3.32 & 20.18 (0.02) & \emph{r} & LT & 09/04/2011& 55660.95 & 30.15 & 20.47 (0.08) & \emph{i} & LT &	 02/06/2011 & 55714.97 & 65.60 & 20.92 (0.07) & \emph{i} & LT \\
28/02/2011& 55620.06 & 3.32 & 20.20 (0.03) & \emph{i} & LT & 10/04/2011 & 55661.94 & 30.80 & 21.36 (0.07) & \emph{g} & LT & 06/06/2011 & 55718.26 & 67.76 & 21.08 (0.05) & \zps & PS1 \\
28/02/2011& 55620.07 & 3.32 & 20.24 (0.06) & \emph{z} & LT & 10/04/2011 & 55661.95 & 30.80 & 20.75 (0.07) & \emph{r} & LT & 07/06/2011 & 55719.27 & 68.42 & 22.71 (0.13) & \gps & PS1 \\
01/03/2011& 55621.02 & 3.95 & 20.49 (0.02) & \emph{g} & LT & 10/04/2011 & 55661.95 & 30.81 & 20.53 (0.05) & \emph{i} & LT & 07/06/2011 & 55719.28 & 68.43 & 21.73 (0.05) & \rps & PS1 \\
01/03/2011& 55621.03 & 3.96 & 20.17 (0.02) & \emph{r} & LT & 10/04/2011 & 55661.96 & 30.81 & 20.60 (0.10) & \emph{z} & LT & 08/06/2011 & 55720.30 & 69.09 & 21.18 (0.07) & \ips & PS1 \\
01/03/2011& 55621.04 & 3.96 & 20.21 (0.02) & \emph{i} & LT & 11/04/2011 & 55662.33 & 31.05 & 21.36 (0.05) & \gps & PS1 & 08/06/2011 & 55720.89 & 69.48 & 22.32 (0.16) & \emph{g} & LT \\
01/03/2011& 55621.05 & 3.97 & 20.41 (0.07) & \emph{z} & LT & 11/04/2011 & 55662.34 & 31.06 & 20.68 (0.03) & \rps & PS1 &  08/06/2011 & 55720.90 & 69.49 & 21.60 (0.07) & \emph{r} & LT \\
09/03/2011& 55629.39 & 9.44 & 20.55 (0.02) & \gps & PS1 & 11/04/2011 & 55662.94 & 31.46 & 21.32 (0.12) & \emph{g} & LT & 08/06/2011 & 55720.91 & 69.50 & 21.01 (0.07) & \emph{i} & LT \\
09/03/2011& 55629.41 & 9.45 & 20.17 (0.02) & \rps & PS1 & 12/04/2011 & 55663.31 & 31.70 & 20.60 (0.02) & \ips & PS1 & 11/06/2011 & 55723.28 & 71.05 & 21.20 (0.04) & \ips & PS1 \\
11/03/2011& 55631.38 & 10.75 & 20.35 (0.03) & \zps & PS1 &	 12/04/2011 & 55663.94 & 32.12 & 20.69 (0.07) & \emph{r} & LT & 14/06/2011 & 55726.26 & 73.01 & 21.26 (0.17) & \yps & PS1 \\
12/03/2011& 55632.50 & 11.48 & 20.59 (0.02) & \gps & PS1 &	 12/04/2011 & 55663.95 & 32.12 & 20.48 (0.06) & \emph{i} & LT & 15/06/2011 & 55727.26 & 73.66 & 21.34 (0.16) & \yps & PS1 \\
12/03/2011& 55632.52 & 11.49 & 20.26 (0.02) & \rps & PS1 &	 13/04/2011 & 55664.94 & 32.77 & 20.69 (0.06) & \emph{r} & LT & 25/06/2011 & 55737.90 & 80.65 & 22.65 (0.10) & \emph{g} & LT \\
13/03/2011& 55633.38 & 12.06 & 20.27 (0.02) & \ips & PS1 &13/04/2011 & 55664.95 & 32.78 & 20.58 (0.06) & \emph{i} & LT & 25/06/2011 & 55737.92 & 80.65 & 21.87 (0.09) & \emph{r} & LT \\
14/03/2011& 55634.32 & 12.68 & 20.41 (0.04) & \zps & PS1 &	13/04/2011 & 55664.96 & 32.78 & 20.45 (0.10) & \emph{z} & LT & 29/06/2011 & 55741.89 & 83.26 & 22.74 (0.14) & \emph{g} & LT \\
15/03/2011& 55635.34 & 13.35 & 20.67 (0.04) & \gps & PS1 &	14/04/2011 & 55665.29 & 33.00 & 21.61 (0.09) & \gps & PS1 & 29/06/2011 & 55741.91 & 83.27 & 21.79 (0.08) & \emph{r} & LT \\
15/03/2011& 55635.36 & 13.36 & 20.25 (0.01) & \rps & PS1 &	14/04/2011 & 55665.30 & 33.01 & 20.82 (0.03) & \rps & PS1 &	 29/06/2011 & 55741.92 & 83.28 & 21.35 (0.06) & \emph{i} & LT \\
16/03/2011& 55636.38 & 14.03 & 20.27 (0.04) & \ips & PS1 &	15/04/2011 & 55666.32 & 33.67 & 20.60 (0.02) & \ips & PS1 &	 25/12/2011 & 55920.10 $\pm$15  & 200.20 $\pm$10 & 23.68 (0.17) & \rps & PS1 \\
17/03/2011& 55637.43 & 14.72 & 20.30 (0.15) & \yps & PS1 &	17/04/2011 & 55668.40 & 35.04 & 20.46 (0.09) & \yps & PS1 & 30/12/2011 & 55925.90 $\pm$15 & 204.00 $\pm$10 & 22.60 (0.14) & \zps & PS1 \\
21/03/2011& 55641.49 & 17.38 & 20.51 (0.08) & \yps & PS1 &	18/04/2011 & 55669.39 & 35.69 & 20.68 (0.10) & \yps & PS1 & 29/01/2012 & 55955.10 $\pm$15 & 223.16 $\pm$10 & 23.45 (0.07) & \ips & PS1 \\
21/03/2011& 55641.96 & 17.69 & 20.75 (0.05) & \emph{g} & LT &	20/04/2011 & 55671.39 & 37.00 & 20.62 (0.09) & \yps & PS1& 01/02/2012  & 55958.60 $\pm$15 & 225.46 $\pm$10 & $>$24.63 & \gps & PS1 \\
21/03/2011& 55641.97 & 17.70 & 20.23 (0.03) & \emph{r} & LT &	21/04/2011 & 55672.35 & 37.63 & 20.64 (0.02) & \ips & PS1  & 01/02/2012 & 55958.60 $\pm$15 & 225.46 $\pm$10 & 23.17 (0.13) & \zps & PS1 \\
21/03/2011& 55641.98 & 17.70 & 20.28 (0.04) & \emph{i} & LT &	22/04/2011 & 55673.34 & 38.28 & 20.73 (0.03) & \zps & PS1 & 21/02/2012 & 55978.50 $\pm$15 & 238.52 $\pm$10 & 23.61 (0.18) & \ips & PS1 \\
21/03/2011& 55641.99 & 17.71 & 20.39 (0.06) & \emph{z} & LT &	22/04/2011 & 55673.89 & 38.64 & 21.69 (0.09) & \emph{g} & LT & 17/03/2012 & 56003.50 $\pm$15 & 254.92 $\pm$10 & $>$24.44 & \rps & PS1 \\
22/03/2011& 55642.95 & 18.34 & 20.81 (0.02) & \emph{g} & LT &	23/04/2011 & 55674.31 & 38.92 & 21.74 (0.04) & \gps & PS1 & 31/03/2012 & 56017.30 $\pm$15 & 263.98 $\pm$10 & 23.62 (0.19) & \ips & PS1 \\
\hline
 \end{tabular} 
 \medskip 
\end{table}
\end{landscape}

\begin{table*}
 \caption{K-correction values for \ap.  The notation $K_{x\to y}$ is
   used where \emph{x} is the observed \PS\ filter and \emph{y} is the
   transformed filter.  The conversation filters correspond to
   standard SDSS filters except for the \emph{UV} which is close to
   uvw1 of Swift. All magnitudes are in the AB system.}
 \label{tab:11ap_kc}
 \begin{tabular}{@{}lcccccc}
  \hline
  \hline
Date & $K_{g\to UV}$ & $K_{r\to u}$ & $K_{i\to g}$ & $K_{z\to r}$ & $K_{y\to i}$ \\
    \hline
02/2011 & 0.27 (0.02) & 0.42 (0.02) & 0.47 (0.02) & 0.48 (0.02) & 0.44 (0.07) \\ 
03/2011 & 0.09 (0.01) & 0.72 (0.02) & 0.40 (0.05) & 0.66 (0.02) & 0.68 (0.08) \\
04/2011 & 0.30 (0.02) & 0.86 (0.09) & 0.49 (0.05) & 0.70 (0.05) & 0.71 (0.10) \\
06/2011 & -0.07 (0.01) & 1.04 (0.08) & 0.57 (0.04) & 0.75 (0.03) & 0.80 (0.09) \\
\hline
 \end{tabular}
 \medskip
\end{table*}

\begin{table*}
\scriptsize
 \caption{The bolometric light curve data for \ap, as seen in Figure\,10.}
 \label{tab:11ap_bollc}
 \begin{tabular}{@{}cc}
  \hline
  \hline
Phase (days, rest) & LogL (ergs$^{-1}$)\\
    \hline
-36.39 & 43.44 $\pm$ 0.046 \\
-30.52 & 43.62 $\pm$ 0.046 \\
-26.54 & 43.73 $\pm$ 0.044 \\
-18.65 & 43.87 $\pm$ 0.040 \\
-14.84 & 43.89 $\pm$ 0.040 \\
-12.83 & 43.91 $\pm$ 0.039 \\
-5.28 & 43.91 $\pm$ 0.043 \\
-1.34 & 43.91 $\pm$ 0.038 \\
-0.68 & 43.92 $\pm$ 0.037 \\
0.00 & 43.92 $\pm$ 0.038 \\
1.29 & 43.90 $\pm$ 0.036 \\
1.96 & 43.91 $\pm$ 0.036 \\
2.60 & 43.91 $\pm$ 0.036 \\
10.71 & 43.86 $\pm$ 0.039 \\
12.68 & 43.85 $\pm$ 0.039 \\
13.37 & 43.85 $\pm$ 0.041 \\
16.34 & 43.84 $\pm$ 0.036 \\
16.99 & 43.81 $\pm$ 0.032 \\
17.68 & 43.81 $\pm$ 0.033 \\
18.98 & 43.80 $\pm$ 0.032 \\
18.98 & 43.79 $\pm$ 0.033 \\
19.61 & 43.78 $\pm$ 0.031 \\
20.36 & 43.78 $\pm$ 0.032 \\
20.94 & 43.78 $\pm$ 0.032 \\
21.60 & 43.75 $\pm$ 0.033 \\
22.58 & 43.75 $\pm$ 0.033 \\
22.89 & 43.74 $\pm$ 0.030 \\
24.86 & 43.66 $\pm$ 0.037 \\
25.51 & 43.65 $\pm$ 0.038 \\
26.18 & 43.65 $\pm$ 0.037 \\
28.14 & 43.61 $\pm$ 0.039 \\
28.79 & 43.60 $\pm$ 0.040 \\
29.45 & 43.59 $\pm$ 0.040 \\
30.35 & 43.59 $\pm$ 0.042 \\
30.76 & 43.59 $\pm$ 0.041 \\
31.42 & 43.56 $\pm$ 0.040 \\
32.32 & 43.53 $\pm$ 0.037 \\
36.28 & 43.50 $\pm$ 0.037 \\
38.23 & 43.49 $\pm$ 0.036 \\
38.60 & 43.50 $\pm$ 0.034 \\
39.30 & 43.48 $\pm$ 0.035 \\
42.13 & 43.40 $\pm$ 0.035 \\
45.16 & 43.42 $\pm$ 0.035 \\
46.48 & 43.43 $\pm$ 0.036 \\
47.79 & 43.40 $\pm$ 0.040 \\
49.36 & 43.37 $\pm$ 0.042 \\
51.76 & 43.36 $\pm$ 0.039 \\
53.70 & 43.36 $\pm$ 0.041 \\
56.32 & 43.37 $\pm$ 0.037 \\
57.68 & 43.38 $\pm$ 0.038 \\
57.87 & 43.36 $\pm$ 0.035 \\
61.83 & 43.32 $\pm$ 0.035 \\
62.88 & 43.33 $\pm$ 0.036 \\
64.23 & 43.34 $\pm$ 0.036 \\
67.74 & 43.27 $\pm$ 0.046 \\
68.13 & 43.31 $\pm$ 0.048 \\
69.70 & 43.28 $\pm$ 0.047 \\
81.91 & 43.19 $\pm$ 0.044 \\
84.51 & 43.20 $\pm$ 0.040 \\
87.13 & 43.19 $\pm$ 0.045 \\
221.81 & 42.61 $\pm$ 0.066 \\
237.16 & 42.54 $\pm$ 0.070 \\
\hline
 \end{tabular}
 \medskip
\end{table*}

\end{document}